\documentclass[11pt, prd, aps, showpacs, nofootinbib]{revtex4-1}

\usepackage{graphicx}
\usepackage{amssymb}
\usepackage{appendix}

\newcommand{\be}{\begin{equation}}
\newcommand{\ee}{\end{equation}}
\newcommand{\bea}{\begin{eqnarray}}
\newcommand{\eea}{\end{eqnarray}}

\newcommand{\Romatre}{Dipartimento di Matematica e Fisica, Universit\`a  degli Studi Roma Tre and INFN, Sezione di Roma Tre,\\ Via della Vasca Navale 84, I-00146 Rome, Italy}
\newcommand{\RomatreINFN}{Istituto Nazionale di Fisica Nucleare, Sezione di Roma Tre,\\ Via della Vasca Navale 84, I-00146 Rome, Italy}

\begin{document}

\title{Extraction of multiple exponential signals from lattice correlation functions}

\author{S.~Romiti} \affiliation{\Romatre}
\author{S.~Simula} \affiliation{\RomatreINFN}

\begin{abstract}
We present a fast and simple algorithm that allows the extraction of multiple exponential signals from the temporal dependence of correlation functions evaluated on the lattice including the statistical fluctuations of each signal and treating properly backward signals.
The method starts from well-known features of the solution of ordinary (linear) differential equations (ODEs) and extracts multiple exponential signals from a generic correlation function simply by inverting appropriate matrices and by finding the roots of an appropriate polynomial, constructed using discretized derivatives of the correlation function.
The method is tested strictly using fake data generated assuming a fixed number of exponential signals included in the correlation function with a controlled numerical precision and within given statistical fluctuations.
All the exponential signals together with their statistical uncertainties are determined exactly by the ODE algorithm.
The only limiting factor is the numerical rounding off.
We show that, even when the total number of exponential signals contained in the correlation function is not known, the ODE method guarantees a quite good convergence toward accurate results for both masses and amplitudes, including their statistical fluctuations, at least for a significant subset of the exponential signals present in the correlation function.
In the case of correlation functions evaluated by large-scale QCD simulations on the lattice various sources of noise, other than the numerical rounding, can affect the correlation function and they represent the crucial factor limiting the number of exponential signals, related to the hadronic spectral decomposition of the correlation function, that can be properly extracted.
The ODE algorithm can be applied to a large variety of correlation functions typically encountered in QCD or QCD+QED simulations on the lattice, including the case of exponential signals corresponding to poles with arbitrary multiplicity and/or the case of oscillating signals. 
Among the appealing features of the ODE algorithm we mention its ability to detect the specific structure of the multiple exponential signals without any {\it a priori} assumption and the possibility to determine accurately the ground-state signal without the need that the lattice temporal extension is large enough to allow the ground-state signal to be isolated.
\end{abstract}

\maketitle

\newpage

\section{Introduction}
\label{sec:intro}

The investigation of the limits of the Standard Model of particle physics and the search for possible signatures of New Physics represent nowadays the main task for both experimental and theoretical physics.
As present (and planned) experimental facilities explore new energy frontiers and improve the precision of the measurements, the importance of flavor physics is continuously growing. 
It is crucial for such studies to quantify the non-perturbative effects due to the strong interaction among hadrons in physical processes.
Large-scale numerical simulations of Quantum Chromodynamics (QCD) performed on the lattice allow for an {\it ab initio} computation of hadronic quantities based on first principles only.
Since several years various quantities relevant for flavour physics phenomenology have been determined by lattice QCD simulations reaching the impressive level of precision of ${\cal{O}}(1 \%)$ or even better (see the recent FLAG-4 review~\cite{Aoki:2019cca}).
Moreover, in recent years lattice computations include also isospin-breaking effects due to electromagnetism and to the mass difference between up and down quarks, which are of the order of ${\cal{O}}(\alpha_{em})$ and ${\cal{O}}[(m_d - m_u) / \Lambda_{\mathrm{QCD}}]$, i.e.~both of the order of ${\cal{O}}(1 \%)$.
Thus, QCD+QED simulations on the lattice are now crucial for making further progresses in flavor physics phenomenology.

The main quantities evaluated through QCD (or QCD+QED) simulations on the lattice are correlation functions defined in terms of the expectation value on the vacuum of the time-ordered product of operators appropriate for the investigation of the physical process of interest.
The operators are located at different sites on the lattice. 
In the case of 2-point correlation functions the above sites are referred to as the source and the sink.
Generally speaking the correlation function is integrated over the spatial extension of the lattice to get its dependence on the time distance $t$ between the source and the sink.
Since lattice simulations are defined in the Euclidean space, the temporal dependence of a correlation function admits a spectral decomposition in terms of a sum of exponential signals of the form $A e^{- M t}$, where $M$ is a hadron mass (or energy when the hadron is moving), corresponding to an eigenvalue of the QCD (or QCD+QED) Hamiltonian, and $A$ is the related amplitude containing a hadronic matrix element, which can be of interest.  

In this work we present a fast and simple algorithm that allows the extraction of multiple exponential signals from the temporal dependence of correlation functions evaluated on the lattice including the determination of the statistical fluctuations of each signal.
The method starts from well-known features of the solution of ordinary (linear) differential equations (ODEs) and extracts multiple exponential signals from a generic correlation function simply by inverting appropriate matrices and by finding the roots of an appropriate polynomial, constructed using discretized derivatives of the correlation function.

The idea of using properties of ODEs for extracting multiple exponential signals from data is not at all a new one and it traces back to the method originally developed by Gaspard Riche de Prony long time ago~\cite{Prony}.
Since then, there have been many developments and applications aimed at fitting data with linear combinations of real (or complex) exponentials within many disciplines in Science and Engineering (see, e.g., Ref.~\cite{PS} and references therein).
In the last decades Prony-type methods~\cite{PS,OS} have been used to determine the masses of excited states from lattice correlators (see Refs.~\cite{Fleming:2004hs,Fleming:2009wb,Beane:2009kya} and more recently Ref.~\cite{Cushman:2019tcv}).
These methods are based on the construction of a difference equation which involves data at equally spaced values in the time direction and the coefficients of the exponents are determined by finding the roots of a characteristic polynomial.
Our method, which we have developed having in mind applications to the analysis of lattice correlation functions, is a Prony-type method characterized by the direct use of discretized derivatives of the correlation function.

The paper is organized as follows.
In Section~\ref{sec:ODE} the basic ingredients of the ODE algorithm are presented, namely the {\it mass} and {\it amplitude} matrices.
The first step is the inversion of the mass matrix, which allows to construct the appropriate polynomial whose roots provide the {\it masses} of the exponential signals (i.e.~the coefficients in the exponent).
Subsequently, the amplitude matrix is constructed and inverted to determine the {\it amplitudes} of the exponential signals.

It will be shown that the ODE algorithm can be applied to a large variety of correlation functions typically encountered in large-scale QCD (or QCD+QED) simulations on the lattice, including the case of exponential signals corresponding to poles with arbitrary multiplicity and/or the case of oscillating signals.
The ODE algorithm is able to detect the specific structure of the exponential signals (single/multiple poles or oscillating signals) without any {\it a priori} knowledge.
This feature is a remarkable one, since it allows to specify the fitting ansatz appropriate for the lattice correlator.

The case of correlators with definite time parity is discussed in Section~\ref{sec:t-parity}, while those corresponding to oscillating signals, poles with arbitrary multiplicity and multiple correlators are addressed in Sections~\ref{sec:staggered},~\ref{sec:multiplicity} and \ref{sec:LS}, respectively.
In Section~\ref{sec:filtering} we briefly discuss the issue of removing unwanted exponential signals from a correlator. 
Thanks to the ODE algorithm it is possible to filter out such signals without the need of determining their amplitudes, as described in details in the Appendix~\ref{sec:appendix}.

In Section~\ref{sec:condition} the main features of the mass and amplitude matrices are discussed. 
Such matrices may be close to singularity and, therefore, the use of an appropriate numerical precision for matrix inversion is crucial for obtaining reliable results.
The closeness to singularity can be described by means of the matrix condition number. 
On one hand side high values of the condition number represent a positive feature, that allows the ODE algorithm to be sensitive to the statistical fluctuations of the exponential signals.
On the other hand side they may be a limiting factor related to the presence of noise in the correlation function, which can be produced by the numerical rounding and/or by other sources.

In Section~\ref{sec:fake_data} the ODE algorithm is tested strictly using {\it fake} data for the correlation function, which are generated assuming a fixed number of exponential signals included in the correlation function and a controlled numerical precision. 
Each exponential signal is allowed to fluctuate within given uncertainties.
We point out that, even if fake correlators represent an ideal situation different from the case of lattice correlators, it is nevertheless mandatory to check that the ODE algorithm (as well as any other algorithm developed for fitting the temporal dependence of lattice correlators) is able to provide exact results in a controlled situation.
This is indeed our case and we show that the ODE algorithm is able to extract exactly all the exponential signals used as input together with their statistical uncertainties.
An important, general feature of the ODE algorithm is that the ground-state signal can be extracted with accuracy even if the lattice temporal extension is not large enough to allow the ground-state signal to be isolated.
This is a very useful property, which in particular can take care properly of the contamination of the excited states in the lattice correlators used for the determination of several hadronic quantities (like, e.g., the form factors).

In Section~\ref{sec:N_ODE} the case of correlators containing more exponential signals than those searched for is discussed.
It will be shown how the ODE algorithm is still able to detect properly several exponential signals. 

In Section~\ref{sec:N_ODE+} the opposite situation, in which more signals than those included in the fake data are searched for, is investigated.
It is found that the noise produced by the numerical rounding generates extra signals, which are not (or only weakly) suppressed exponentially in the time distance. 

In Section~\ref{sec:lattice} the ODE algorithm is applied to the case of {\it real} data, namely correlation functions evaluated by means of large-scale QCD simulations on the lattice. 
Various sources of noise, other than the numerical rounding, can now affect the correlator, like e.g.~the residues coming from gauge variant terms.
It will be shown that the noise becomes the crucial factor limiting the number of exponentials, related to the eigenstates of the QCD Hamiltonian, that can be extracted from the correlation function.

In Section~\ref{sec:improvement} the combination of the ODE algorithm with other techniques suitable for fitting lattice correlators is briefly discussed.
An interesting possibility is represented by the combination of the ODE algorithm with a subsequent nonlinear least-squares minimizer, where masses and amplitudes are used as free parameters to be varied (without any prior) starting from the values obtained by the ODE method.
In the case of the lattice correlators analyzed in Section~\ref{sec:lattice} the ODE solution is nicely confirmed, within the uncertainties, by the subsequent $\chi^2$-minimization procedure.

Finally, Section~\ref{sec:conclusions} contains our conclusions.

\section{The ODE method}
\label{sec:ODE}

Let's start by considering a correlator $C(t)$ composed by $N^{(+)}$ exponential signals in the forward time direction and $N^{(-)}$ exponentials in the backward one:
\be
     C(t) = \sum_{i=1}^{N^{(+)}} A_i^{(+)} e^{- M_i^{(+)} t} + \sum_{j=1}^{N^{(-)}} A_j^{(-)} e^{- M_j^{(-)} (T - t)} ,
     \label{eq:Ct}
\ee
where $T$ is the temporal extension of the lattice.
In Eq.~(\ref{eq:Ct}) the {\it masses} $M_i^{(+)}$ and $M_j^{(-)}$ are nonnegative real numbers (as in the case of hadronic masses) and the {\it amplitudes} $A_i^{(+)}$ and $A_j^{(-)}$ are real numbers\footnote{The generalization to the case of complex amplitudes is straightforward by adopting a strategy similar to the one described in Section~\ref{sec:LS}.}.

The correlator $C(t)$ is supposed to be known at discretized values of the time distance $t$, namely $t \equiv n a$ with $n = 1, ... \, N_T$, where $a$ is the lattice spacing and $N_T \equiv T /a$ is the number of lattice points in the temporal direction.
In lattice QCD (or QCD+QED) simulations a correlator of the form (\ref{eq:Ct}) may correspond, e.g., to the case of a nucleonic correlator, where the backward signals correspond to negative parity partners of the nucleon and its excitations.

For sake of simplicity we will refer to the quantities $M_i^{(+)}$ and $M_j^{(-)}$ as masses. 
It is however clear that Eq.~(\ref{eq:Ct}) may correspond also to the case of correlation functions for moving hadrons by simply replacing hadron masses with energies. 

The correlator (\ref{eq:Ct}) can be rewritten as
\be
     C(t = n a) \equiv C_n^{(0)} = \sum_{m=1}^N \widetilde{A}_m e^{- a \widetilde{M}_m n}
     \label{eq:C0n}
\ee
with $N \equiv N^{(+)}+ N^{(-)}$ and
\be
    \widetilde{M}_m = M_i^{(+)} ~, ~ \qquad \qquad \widetilde{A}_m = A_i^{(+)}
    \label{eq:forward_MA}
\ee
in the case of forward signals ($m = i = 1, ... \, N^{(+)} $)
and 
\be
    \widetilde{M}_m = - M_j^{(-)}  ~, ~ \qquad \qquad \widetilde{A}_m = A_j^{(-)}e^{-M_j^{(-)} T}
    \label{eq:backward_MA}
\ee
for backward signals ($m = N^{(+)} + j = N^{(+)} +1, ... \, N^{(+)} + N^{(-)}$).

Let's now consider the discretized (symmetric) time derivative
\be
    \label{eq:C1n}
     C_n^{(1)} \equiv \frac{1}{2} \left[ C_{n+1}^{(0)} - C_{n-1}^{(0)} \right] =
                                \sum_{m=1}^N \widetilde{A}_m \, z_m \, e^{- a \widetilde{M}_m n} ~ , 
\ee
where
\be
    z_m \equiv - \mbox{sinh}(a \widetilde{M}_m) ~ .
    \label{eq:roots}
\ee
By repeating the application of the differential operation (\ref{eq:C1n}) one gets
\be
    \label{eq:Ckn}
    C_n^{(k)} = \frac{1}{2} \left[ C_{n+1}^{(k-1)} - C_{n-1}^{(k-1)} \right] =
                       \sum_{m=1}^N \widetilde{A}_m (z_m)^k e^{- a \widetilde{M}_m n} ~ . 
\ee

We also assume that the correlator $C_n^{(0)}$ is known, for each time distance, in terms of a number of jackknife or bootstrap events. 
Its statistical error $\sigma_n^{(0)}$ can be correspondingly evaluated. 
Starting from the correlator $C_n^{(0)}$, the sequence of the correlators $C_n^{(k)}$ with $k = 1, ... \, N$ can be evaluated  for each jackknife or bootstrap event together with their statistical errors $\sigma_n^{(k)}$.
It is understood that what follows applies for each single jackknife or bootstrap event. 

The values of the correlator $C_n^{(0)}$ are provided in the range $n = [1, \, N_T]$, while the derivatives $C_n^{(k)}$ for $k = 1, ... \, N$ are evaluated only in the range $n = [k+1, \, N_T - k]$.
Outside this range the values of the derivatives are not independent and, therefore, we put $C_n^{(k)} = 0$.

Note that, because of the presence of the factor $(z_m)^k$ in Eq.~(\ref{eq:Ckn}), at a fixed value of $n$ the impact of the signals with higher masses increase as the order $k$ of the derivative $C_n^{(k)}$ increases.

The central step in our procedure consists in introducing $N+1$ real coefficients $x_k$ ($k = 0, 1, ...\, N$) and considering the quantity
\be
    \sum_{k=0}^N x_k C_n^{(k)} = \sum_{m=1}^N \widetilde{A}_m \left[ \sum_{k=0}^N x_k z_m^k \right] e^{- a \widetilde{M}_m n} 
                                                 =  \sum_{m=1}^N \widetilde{A}_m P_N(z_m) e^{- a \widetilde{M}_m n} ~ ,
\ee
where the polynomial $P_N(z)$ of degree N is given by
\be
    P_N(z) \equiv  \sum_{k=0}^N x_k z^k ~ .
    \label{eq:PNz_xk}
\ee
The above polynomial has in general $N$ roots depending on the coefficients $x_k$.
If the latter ones (which, we stress, are independent of $n$) are chosen so that the polynomial $P_N(z)$ has its roots at $z = z_m$ given by Eq.~(\ref{eq:roots}), then the condition
\be
    \sum_{k=0}^N x_k C_n^{(k)} = 0 
    \label{eq:xk}
\ee
holds for any value of $n$.
Note that the roots $z_m$ of the polynomial $P_N(z)$ depend only on the masses $\widetilde{M}_m$ and are independent of the amplitudes $\widetilde{A}_m$.
Moreover, from Eq.~(\ref{eq:roots}) it follows that the roots $z_m$ are real numbers, positive for backward signals and negative for forward ones.

Equation~(\ref{eq:xk}) is a typical ordinary (linear) differential equation (ODE).
Usually the coefficients $x_k$ are given and, therefore, the solution of Eq.~(\ref{eq:xk}) corresponds to Eq.~(\ref{eq:C0n}) with the masses $\widetilde{M}_m$ given by the roots (\ref{eq:roots}) of the polynomial $P_N(z)$ and with the amplitudes $\widetilde{A}_m$ depending on a suitable number of initial conditions.

Here we are interested in the inverse problem: starting from the known values of the correlator $C_n^{(0)}$and its derivatives we want to determine the coefficients $x_k$ of the polynomial (\ref{eq:PNz_xk}) having its roots at $z = z_m$.
The procedure, which hereafter will be referred to as the ODE algorithm, is as follows.

Without any loss of generality we can put $x_N = 1$ so that
\be
    P_N(z) \equiv  \sum_{k=0}^N x_k z^k = \sum_{k=0}^{N-1} x_k z^k + z^N = \prod_{m=1}^N (z - z_m)
    \label{eq:PNz}
\ee
and Eq.~(\ref{eq:xk}) can be rewritten as
\be
     \sum_{k=0}^{N-1} x_k C_n^{(k)} = - C_n^{(N)} ~ . 
     \label{eq:xk_ODE}
\ee
The problem is to solve Eq.~(\ref{eq:xk_ODE}) for the $N$ unknowns $x_k$ ($k=0, 1, ..., N-1$). 

Our aim is to extract the multiple exponential signals in the correlator (\ref{eq:C0n}) using as input the knowledge of the correlator in a given range of values of $n$, which eventually can span the full temporal extension $[1, \, N_T]$.
Therefore, we multiply Eq.~(\ref{eq:xk_ODE}) by a set of $N$ functions $R_n^{(k^\prime)}$ with $k^\prime = 0, 1, ... \, (N - 1)$ and sum over $n$ in a given range from $n_{min}$ to $n_{max}$.
Since the largest range in which the derivatives $C_n^{(k)}$ can be calculated is $n = [k+1, \, N_T - k]$, we put directly $n_{min} = N+1$ and $n_{max} = N_T - N$, so that all the values of the correlator (\ref{eq:C0n}) in the full range $[1, \, N_T]$ are taken into account.
We get the following system of inhomogeneous linear equations
\be
     \sum_{k=0}^{N-1} M_{k^\prime k} x_k  = V_{k^\prime} ~ ,
     \label{eq:mass_eq}
\ee
where the $N \times N$ {\it mass} matrix $M$ is given by
\be
      M_{k^\prime k} \equiv  \sum_{n = N + 1}^{N_T - N} R_n^{(k^\prime)} \, C_n^{(k)} ~ ,
      \label{eq:mass_matrix_generic}
\ee
and the vector $V$ with dimension $N$ by
\be
    V_{k^\prime} \equiv - \ \sum_{n = N + 1}^{N_T - N} R_n^{(k^\prime)} \, C_n^{(N)} ~ .
    \label{eq:vectorV_generic}
\ee

The choice of the functions $R_n^{(k^\prime)}$ is in principle arbitrary, provided it leads to a non-singular mass matrix $M$.
We have explored different choices for $R_n^{(k^\prime)}$.
Simple and natural choices are either
\be
    R_n^{(k^\prime)} = \frac{C_n^{(k^\prime)}}{[\sigma_n^{(0)}]^2} ~ ,
    \label{eq:Rn}
\ee
where $\sigma_n^{(0)}$ is the uncertainty of the correlator (\ref{eq:C0n}), or
\be
    R_n^{(k^\prime)} = \frac{C_n^{(k^\prime)}}{[\sigma_n^{(k^\prime)}]^2} ~ ,
    \label{eq:Rn_alt}
\ee
where $\sigma_n^{(k^\prime)}$ is the uncertainty of the derivative (\ref{eq:Ckn}).
A more sophisticated choice is 
\be
    R_n^{(k^\prime)} = D_{n n^\prime}^{(0)} C_{n^\prime}^{(k^\prime)} ~ ,
    \label{eq:Rn_cov}
\ee
where $D_{n n^\prime}^{(0)}$ is the inverse of the covariance matrix of the correlator (\ref{eq:C0n}).
We have checked that the performance of the ODE algorithm is not changed by the three choices (\ref{eq:Rn}-\ref{eq:Rn_cov}) (see later Section~\ref{sec:condition}).
In what follows we will use the definition (\ref{eq:Rn}).
We point out that any autocorrelation between different values of $n$ is taken into account by the use of the jackknife (bootstrap) procedure.

Thus, we rewrite Eqs.~(\ref{eq:mass_matrix_generic}) and (\ref{eq:vectorV_generic}) as 
\be
      M_{k^\prime k} \equiv \sum_{n = N + 1}^{N_T - N} \frac{C_n^{(k^\prime)} \, C_n^{(k)}}{[\sigma_n^{(0)}]^2} ~ ,
      \label{eq:mass_matrix}
\ee
and 
\be
    V_{k^\prime} \equiv - \sum_{n = N + 1}^{N_T - N} \frac{C_n^{(k^\prime)} \, C_n^{(N)}}{[\sigma_n^{(0)}]^2} ~ .
    \label{eq:vectorV}
\ee
Equations (\ref{eq:mass_matrix}) and (\ref{eq:vectorV}) are evaluated starting from the correlator (\ref{eq:C0n}) and its derivatives (\ref{eq:Ckn}) for each jackknife or bootstrap event.
Note that using the definition (\ref{eq:Rn}) for $R_n^{(k^\prime)}$ the system of linear equations~(\ref{eq:mass_eq}) corresponds to minimize the variable $\chi_M^2$ defined as
\be
    \chi_M^2 \equiv \sum_{n = N + 1}^{N_T - N} \frac{1}{[\sigma_n^{(0)}]^2} \left[ \sum_{k^\prime=0}^N x_{k\prime} \, C_n^{(k^\prime)} \right]^2 ~ ,
    \label{eq:chi2_mass}
\ee
i.e.~to the constraints $\partial \chi_M^2 / \partial x_k = 0$ for $k=0, 1, ... (N-1)$ with $x_N = 1$.

For a non-singular matrix $M$ the coefficients $x_k$ can be determined by inverting the matrix $M$:
\be
    x_k = \sum_{k^\prime=0}^{N-1} M_{k k^\prime}^{-1} \, V_{k^\prime} ~ .
    \label{eq:xk_eq}
\ee
Once the coefficients $x_k$ are known, the roots of the polynomial $P_N(z)$ can be calculated, and thus the masses of the forward and backward exponential signals in lattice units, $aM_i^{(+)}$ and $aM_j^{(-)}$, can be determined from Eq.~(\ref{eq:roots}).

The last step is the determination of the amplitudes $\widetilde{A}_m$.
To this end we introduce a $\chi^2$-variable defined as
\bea
    \label{eq:chi2_N}
    \chi^2 & \equiv & \sum_{k=0}^N \chi_k^2 ~ , \\[2mm]
    \label{eq:chi2_k}
    \chi_k^2 & \equiv & \sum_{n=k+1}^{N_T-k} \left( \frac{C_n^{(k)} - \sum_{m=1}^N \widetilde{A}_m z_m^k 
                                   e^{- a \widetilde{M}_m n}}{\sigma_n^{(k)}} \right)^2 ~ ,
\eea
where $\sigma_n^{(k)}$ is the statistical error of the derivative $C_n^{(k)}$.
Then, we impose the minimization condition $\partial \chi^2 / \partial \widetilde{A}_m = 0$, which leads to the following linear system of equations
\be
    \sum_{m^\prime=1}^N A_{m m^\prime} \widetilde{A}_{m^\prime}  = W_m ~ ,
     \label{eq:amplitude_eq} 
\ee
where
\bea
     \label{eq:amplitude_matrix}
     A_{m m^\prime} & \equiv & \sum_{k=0}^N \sum_{n=k+1}^{N_T-k} ( z_m z_{m^\prime} )^k ~ 
                                                 \frac{e^{- a (\widetilde{M}_m +\widetilde{M}_{m^\prime}) n}}{[\sigma_n^{(k)}]^2} ~ , \\[2mm]
     \label{eq:vectorW}
     W_m & \equiv & \sum_{k=0}^N \sum_{n=k+1}^{N_T-k} (z_m)^k ~ \frac{e^{- a \widetilde{M}_m n} \, C_n^{(k)}}{[\sigma_n^{(k)}]^2} ~ .
\eea
The solution of the linear equation (\ref{eq:amplitude_eq}) is given by
\be
    \widetilde{A}_m = \sum_{m^\prime=1}^N A_{m m^\prime}^{-1} \, W_{m^\prime} ~ ,
    \label{eq:Am_eq}
\ee
which allows to extract the forward and backward amplitudes, $A_i^{(+)}$ and $A_j^{(-)}$ using Eqs.~(\ref{eq:forward_MA}-\ref{eq:backward_MA}).

We stress that the choice of the $\chi^2$-variable given by Eqs.~(\ref{eq:chi2_N}-\ref{eq:chi2_k}) is not unique and, consequently, also the definitions of the amplitude matrix (\ref{eq:amplitude_matrix}) and vector (\ref{eq:vectorW}).
For instance, one can limit the sum in Eq.~(\ref{eq:chi2_N}) to the first term $k = 0$ without involving the derivatives $C_n^{(k)}$ with $k > 0$.
Correspondingly also in Eqs.~(\ref{eq:amplitude_matrix}-\ref{eq:vectorW}) the sum over $k$ should be limited to the first term $k = 0$ only.
Later in Sections~\ref{sec:condition} and~\ref{sec:lattice} we will check explicitly that the changes in the calculated amplitudes due to different choices of the $\chi^2$-variable (\ref{eq:chi2_N}) are totally negligible. 

The key feature of the ODE method is the inversion of the mass and amplitude matrices, given respectively by Eqs.~(\ref{eq:mass_matrix}) and (\ref{eq:amplitude_matrix}).
This important issue will be discussed in Section \ref{sec:condition}.
In the following subsections we want to illustrate how the ODE method can be applied to specific forms of the correlation functions typically encountered in QCD (or QCD+QED) simulations on the lattice.

\subsection{Correlators with definite time parity}
\label{sec:t-parity}

Let's consider the case in which the backward signals in the correlator (\ref{eq:Ct}) have the same masses and amplitudes (in absolute value) of those appearing in the forward signals, namely 
\be
     C(t) = \sum_{i=1}^N A_i \left[ e^{- M_i t} + (-)^{p_i} e^{- M_i (T - t)} \right] ~ ,
     \label{eq:Ct_bf}
\ee
where $(-)^{p_i} = \pm 1$ is the parity with respect to the substitution $t \to (T - t)$.
In what follows we will refer to $(-)^{p_i}$ as the $t$-parity of the $i$-th exponential signal.
Note that, because of periodicity, each exponential signal in Eq.~(\ref{eq:Ct_bf}) receives contributions also from multiple wrappings around the lattice time extension. 
The latter ones correspond to a modification of the amplitudes $A_i$ by a multiplicative factor $(1 - e^{-M_i T})^{-1}$, which we consider already absorbed into the amplitude $A_i$ appearing in Eq.~(\ref{eq:Ct_bf}).
The multiplicative factor is numerically small for enough large values of $M_i T$, as it happens in many applications of lattice QCD simulations.

We can easily construct two combinations which have a definite $t$-parity:
\be
    C^{(\pm)}(t) =  \frac{1}{2} \left[ C(t) \pm C(T-t) \right] = \sum_{i=1}^N A_i \left[ e^{- M_i t} \pm e^{- M_i (T - t)} \right] ~ .
    \label{eq:Ct_pm}
\ee
Thus, without any loss of generality we can consider a correlator $C(t)$ of the following form
\be
     C(t) = \sum_{i=1}^N A_i \left[ e^{- M_i t} + (-)^p e^{- M_i (T - t)} \right] ~ ,
     \label{eq:Ct_p}
\ee
where now $(-)^p$ can be equal to either $-1$ or $+1$ for all signals and $N$ stands for the number of independent exponential signals, including now both the forward and the backward parts.

According to the results of the previous Section, the construction of the matrices (\ref{eq:mass_matrix}) and (\ref{eq:amplitude_matrix}) would imply working with matrices of dimension $2N \times 2N$.
Instead, in the case of correlators with definite $t$-parity we can work with matrices having dimension $N \times N$. 
To do that we modify the definition of the derivatives $C_n^{(k)}$.
We now consider only even discretized derivatives and introduce the second derivative of the correlator (\ref{eq:Ct_p}) as
\be
    \label{eq:C2n}
    C_n^{(2)} \equiv C_{n+1}^{(0)} + C_{n-1}^{(0)} - 2 C_n^{(0)} =
                               \sum_{i=1}^N A_i \, \overline{z}_i \left[ e^{- aM_i n} + (-)^p e^{- aM_i (N_T - n)} \right] ~ , ~ 
\ee
where
\be
     \overline{z}_i \equiv 2 \left[ \mbox{cosh}(a M_i) - 1 \right] = \left[ 2 \, \mbox{sinh}\left( \frac{a M_i}{2} \right) \right]^2~ .
     \label{eq:zi}
\ee
By repeating the application of the differential operation (\ref{eq:C2n}) one gets\footnote{We remind that the values of the correlator $C_n^{(0)}$ are provided in the range $n = [1, \, N_T]$, while the derivatives $C_n^{(2k)}$ for $k = 1, .. \, N$ can be determined only in the range $n = [k+1, \, N_T - k]$. Outside this range we consider that $C_n^{(2k)} = 0$.}
\be
    \label{eq:C2kn}
    C_n^{(2k)} = C_{n+1}^{(2k-2)} + C_{n-1}^{(2k-2)} - 2 C_n^{(2k-2)} =
                          \sum_{i=1}^N A_i \, (\overline{z}_i)^k \left[ e^{- aM_i n} + (-)^p e^{- aM_i (N_T - n)} \right] ~ . 
\ee

We now repeat the same steps followed previously in the case of different forward and backward signals.
We write again explicitly the definitions of the mass and amplitudes matrices (and related vectors), since in the next Sections they will be repeatedly mentioned.

The constraint (\ref{eq:xk}) is now replaced by
\be
     \sum_{k=0}^N \overline{x}_k C_n^{(2k)} = 0 ~ ,
     \label{eq:xk_p}
\ee
where the coefficients $\overline{x}_k$ for $k = 0, 1, ..., (N-1)$ can be determined by solving the linear system of equations (choosing $\overline{x}_N = 1$)
\be
     \sum_{k=0}^{N-1} \overline{M}_{k^\prime k} \overline{x}_k  = \overline{V}_{k^\prime} 
     \label{eq:mass_eq_p}
\ee
with the $N \times N$ mass matrix given by
\be
      \overline{M}_{k^\prime k} \equiv \sum_{n = N + 1}^{N_T - N} \frac{C_n^{(2k^\prime)} \, C_n^{(2k)}}{[\sigma_n^{(0)}]^2}
      \label{eq:mass_matrix_p}
\ee
and the $N$-dimensional vector $\overline{V}$ defined as
\be
    \overline{V}_{k^\prime} \equiv - \sum_{n = N + 1}^{N_T - N} \frac{C_n^{(2k^\prime)} \, C_n^{(2N)}}{[\sigma_n^{(0)}]^2} ~ .
    \label{eq:vectorV_p}
\ee
By means of the coefficients $\overline{x}_k$ we can construct the polynomial 
\be
    \overline{P}_N(z) =  \sum_{k=0}^{N-1} \overline{x}_k z^k + z^N = \prod_{i=1}^N (z - \overline{z}_i) ~ ,
     \label{eq:PNz_p}
\ee
which has its roots at $z = \overline{z}_i$ given by Eq.~(\ref{eq:zi}).
Note that the roots $\overline{z}_i$: ~ i) are positive for real values of $M_i$, ~ ii) do not depend on the amplitudes $A_i$ (i.e., the determination of the nonlinear unknowns $M_i$ is independent on the values of the linear ones $A_i$), and ~ iii) are independent of the specific $t$-parity of the correlator (i.e., on the value of $(-)^p$).
Note also that, thanks to the use of even derivatives only [see the r.h.s.~of Eq.~(\ref{eq:C2kn})], the constraint (\ref{eq:xk_p}) can be satisfied simultaneously by the forward and backward parts of each signal.

Following the same procedure described in the previous Section [see Eqs.~(\ref{eq:chi2_N}-\ref{eq:vectorW})] the amplitudes $A_i$ can be obtained by solving the linear system of equations
\be
    \sum_{j=1}^N \overline{A}_{i j} A_j = \overline{W}_i ~ ,
     \label{eq:amplitude_eq_p} 
\ee
where
\bea
     \label{eq:amplitude_matrix_p}
     \overline{A}_{i j} & \equiv & \sum_{k=0}^N \sum_{n=k+1}^{N_T-k} (\overline{z}_i \overline{z}_j)^k \, \frac{f_n^{(i)} \, f_n^{(j)}}{[\sigma_n^{(2k)}]^2} ~ , \\[2mm]
     \label{eq:vectorW_p}
     \overline{W}_i & \equiv & \sum_{k=0}^N \sum_{n=k+1}^{N_T-k} (\overline{z}_i)^k \, \frac{f_n^{(i)} \, C_n^{(2k)}}{[\sigma_n^{(2k)}]^2} ~ 
\eea
with $\sigma_n^{(2k)}$ being the statistical error of the (even) derivative (\ref{eq:C2kn}) and
\be
    f_n^{(i)} \equiv e^{- aM_i n} + (-)^p e^{- aM_i (N_T - n)} ~ .
    \label{eq:fn_i}
\ee
Note that the above functions depend on the specific $t$-parity of the correlator (\ref{eq:Ct_p}).

Before closing this subsection, we remind that at large time distances the signal of the lightest mass (e.g.~$M_1$) is expected to dominate.
Correspondingly the ratio $C_n^{(2)} / C_n^{(0)}$ may exhibit a plateau related to the value of $a M_1$, namely
\be
    \frac{C_n^{(2)}}{C_n^{(0)}} ~ _{\overrightarrow{n \gg 1, (N_T - n) \gg 1}} ~ \overline{z}_1 = 2 \left[ \mbox{cosh}(a M_1) - 1 \right] ~ .
    \label{eq:Meff}
\ee
We can therefore define an {\it effective} mass $M_{eff}^{(cosh)}$ as
\be
    a M_{eff}^{(cosh)} \equiv \mbox{cosh}^{-1} \left[ 1 + \frac{C_n^{(2)}}{2 C_n^{(0)}} \right] ~ _{\overrightarrow{n \gg 1, (N_T - n) \gg 1}} ~ a M_1.
    \label{eq:Meff_cosh}
\ee

\subsection{Oscillating signals}
\label{sec:staggered}

Specific formulations of the QCD action on the lattice may lead to quite specific features of the correlation functions constructed using quark and gluon propagators.
One of such cases is represented by the staggered formulation, in which the correlation functions may have oscillating terms related to the presence of opposite (spatial) parity partners.

In order to simplify the notations we limit ourselves to the case of a correlator with positive $t$-parity, namely
\bea
     \label{eq:Ct_staggered}
     C(t) & = & C^{(+)}(t) + C^{(-)}(t) ~ , \\[2mm]
     \label{eq:Ct_normal}
     C^{(+)}(t) & = & \sum_{i=1}^{N^{(+)}} A_i^{(+)} \left[ e^{- aM_i^{(+)} n} + e^{- aM_i^{(+)} (N_T - n)} \right] ~ , \\[2mm]
     \label{eq:Ct_oscillating}
     C^{(-)}(t) & = & \sum_{j=1}^{N^{(-)}} A_j^{(-)} \left[ (-)^n e^{- aM_j^{(-)} n} + (-)^{N_T-n} e^{- aM_j^{(-)} (N_T - n)} \right] ~ ,
\eea
where now $N^{(\pm)}$ stands for normal/oscillating exponential signals.

The possible presence of oscillating signals do not represent a problem for the ODE algorithm.
Indeed, since $(-)^n = e^{\pm i \pi n}$ the correlator (\ref{eq:Ct_staggered}) can be written as the sum of $N = N^{(+)} + N^{(-)}$ exponential signals as
\be
     C(t = an) \to C_n^{(0)} = \sum_{m=1}^N \widetilde{A}_m \left[ e^{- a \widetilde{M}_m n} + e^{- a \widetilde{M}_m (N_T - n)} \right] ~ ,
     \label{eq:C0n_staggered}
\ee
where
\be
    a\widetilde{M}_m = aM_i^{(+)} ~, ~ \qquad \qquad \qquad \widetilde{A}_m = A_i^{(+)}
    \label{eq:normal_MA}
\ee
in the case of the normal signals ($m = i = 1, ... \, N^{(+)} $)
and 
\be
    a\widetilde{M}_m = aM_j^{(-)}  - i \pi ~, ~ \qquad \qquad \widetilde{A}_m = A_j^{(-)}
    \label{eq:oscillating_MA}
\ee
for oscillating signals ($m = N^{(+)} + j = N^{(+)} +1, ... \, N^{(+)} + N^{(-)}$), for which the masses $a\widetilde{M}_m$ acquire an imaginary part equal to $-\pi$ (or equivalently $+\pi$).  

We can therefore apply our ODE algorithm.
The only difference is that now the roots $\overline{z}_m$, while remaining real, can be either positive or negative.
The positive ones correspond to normal signals, while the negative roots to oscillating signals (more precisely $\overline{z}_m \leq -4$ because $2 \left[ \mbox{cosh}(a M - i \pi) - 1 \right] = -2 \left[ \mbox{cosh}(a M) + 1 \right]$).

Before closing this subsection, we point out that the possibility to detect the presence of oscillating signals occurs when the analyzed correlator has a definite time parity, i.e.~when the relation between masses and roots is given by Eq.~(\ref{eq:zi}), which originates from the use of even derivatives only.
On the contrary, when both odd and even derivatives are used, the relation between masses and roots is given by Eq.~(\ref{eq:roots}).
Since $\mbox{sinh}(aM_i \pm i \pi) = - \mbox{sinh}(aM_i) = \mbox{sinh}(-aM_i)$, a forward(backward) oscillating signal cannot be distinguished from a backward(forward) non-oscillating signal, basing only on the analyses of the mass matrix (\ref{eq:mass_matrix_generic}) constructed using both odd and even derivatives.

\subsection{Poles with arbitrary multiplicity}
\label{sec:multiplicity}

Till now we have considered multiple exponential signals corresponding to single poles of the form $(p^2 - M_i^2)^{-1}$ in (Minkowskian) momentum space.
In this subsection we address the case of poles characterized by an arbitrary multiplicity $\mu_i$, which correspond to exponential signals multiplied by a polynomial of degree ($\mu_i - 1$) in the time distance $t$.
Thus, the correlator (\ref{eq:Ct_p}) is replaced by
\be
     C(t) = \sum_{i=1}^{\overline{N}} \sum_{\mu=0}^{\mu_i-1} B_{i \mu} \left[ t^\mu e^{-M_i t} + (-)^p (T - t)^\mu e^{-M_i (T - t)} \right] ~ ,
     \label{eq:Ct_mu}
\ee
where $N \equiv \sum_{i=1}^{\overline{N}} \mu_i$ represents the total number of exponential terms [i.e., those in the square brackets in the r.h.s.~of Eq.~(\ref{eq:Ct_mu})].
In modern lattice QCD+QED simulations the above situation may occur, e.g., when isospin breaking effects, due to the quark electric charges and to the mass difference $\delta m$ between $u$ and $d$ quarks, are taken into account at leading order in the electromagnetic coupling $\alpha_{em}$ and in $\delta m$ (see, e.g., Ref.~\cite{RM123}).
In this case correlation functions contain double poles (i.e.~$\mu_i = 2$).

Within the ODE algorithm the procedure is as follows.
Let's start from the correlator
\be
     C_n^{(0)} = \sum_{i=1}^{\overline{N}} \sum_{\mu=0}^{\mu_i-1} \overline{B}_{i \mu} \, f_n^{(i, \, \mu - 0)} ~ ,
     \label{eq:C0n_mu}
\ee
where the amplitudes $\overline{B}_{i \mu} \equiv B_{i \mu} a^\mu$ are given in lattice units and
\be
    f_n^{(i, \, \mu - j)} \equiv n^{\mu - j} e^{-aM_i n} + (-)^p (N_T - n)^{\mu -j} e^{-aM_i (N_T - n)} ~ .
    \label{eq:fn_imu}
\ee
The sequence of even derivatives
\be
    C_n^{(2k)} = C_{n+1}^{(2k-2)} + C_{n-1}^{(2k-2)} - 2 C_n^{(2k-2)} 
    \label{eq:C2kn_mu}
\ee
can be constructed for $k = 1, ..., N$.
One gets
\bea
     \sum_{k=0}^N \overline{x}_k C_n^{(2k)} & = & \sum_{i=1}^{\overline{N}}  \overline{P}_N(\overline{z}_i) \sum_{\mu=0}^{\mu_i-1} 
                                                                               \overline{B}_{i \mu} \, f_n^{(i, \, \mu - 0)} \\
                                                                     & + & \sum_{i=1}^{\overline{N}} \frac{d \overline{P}_N}{dz}(\overline{z}_i)  \sum_{\mu=0}^{\mu_i-1}
                                                                               \overline{B}_{i \mu} \sum_{j=1}^\mu {{\mu}\choose{j}} \left[ e^{-aM_i} + (-)^j e^{aM_i} \right]
                                                                               f_n^{(i, \, \mu-j)} \nonumber \\
                                                                    & + & \sum_{i=1}^{\overline{N}} \frac{1}{2!} \frac{d^2 \overline{P}_N}{dz^2}(\overline{z}_i) 
                                                                              \sum_{\mu=0}^{\mu_i-1}  \overline{B}_{i \mu} \sum_{j=1}^\mu {{\mu}\choose{j}} 
                                                                              \left[ e^{-aM_i} + (-)^j e^{aM_i} \right] \nonumber \\ 
                                                                    & \cdot & \sum_{j^\prime=1}^{\mu-j} {{\mu-j}\choose{j^\prime}} \left[ e^{-aM_i} + (-)^{j^\prime} e^{aM_i} 
                                                                                    \right] f_n^{(i, \, \mu-j-j^\prime)} + ... \qquad \nonumber
\eea
where $\overline{P}_N(z) \equiv \sum_{k=0}^N \overline{x}_k z^k$ and $\overline{z}_i$ is given by Eq.~(\ref{eq:zi}).
The constraint $\sum_{k=0}^N \overline{x}_k C_n^{(2k)} = 0$ is satisfied for each value of $n$ only if
\be
    \overline{P}_N(\overline{z}_i) = \frac{d\overline{P}_N}{dz}(\overline{z}_i) = ... = \frac{d^{\mu_i - 1} \overline{P}_N}{dz^{\mu_i - 1}}(\overline{z}_i) = 0 ~ ,
    \label{eq:constraints}
\ee
which means that the root $\overline{z}_i$ of the polynomial $\overline{P}_N(z)$ has multiplicity $\mu_i$, namely (with $\overline{x}_N = 1$)
\be
    \overline{P}_N(z) = \prod_{i=1}^{\overline{N}} (z - \overline{z}_i)^{\mu_i} ~ .
    \label{eq:PNz_mu}
\ee
Thus, the masses $M_i$ and their multiplicities $\mu_i$ ($i=1, 2, ..., M$) are determined in two steps: first by solving the linear system of equations corresponding to Eqs.~(\ref{eq:mass_eq_p}-\ref{eq:vectorV_p}) to obtain the coefficients $x_k$ and then by finding the roots of the polynomial $\overline{P}_N(z) \equiv \sum_{k=0}^{N-1} \overline{x}_k z^k + z^N$ with their multiplicities $\mu_i$.
Finally, the amplitudes $\overline{B}_{i \mu}$ can be determined by solving the linear system of equations corresponding to Eqs.~(\ref{eq:amplitude_eq_p}-\ref{eq:vectorW_p}) with the functions $f_n^{(i)}$ replaced by $f_n^{(i, \mu -0)}$ given in Eq.~(\ref{eq:fn_imu}).

\subsection{Multiple correlators}
\label{sec:LS}

In this subsection we address briefly the case in which different correlators sharing the same masses $M_i$ are available.
In lattice QCD (or QCD+QED) simulations such a situation may occur when different interpolating fields, like e.g.~local (L) and smeared (S) fields, are adopted in the source and in the sink.

For sake of simplicity let's consider the simple case of four correlators $C^{f f^\prime}(t)$ with $f(f^\prime) = L, S$.
We look for multiple exponential signals of the form
\be
     C^{f f^\prime}(t) \to \sum_{i=1}^N A_i^f A_i^{f^\prime} \left[ e^{- M_i t} + (-)^p e^{- M_i (T - t)} \right] ~.
     \label{eq:Ct_multiple}
\ee
We assume also that for each jackknife (or bootstrap) event one has $C^{LS}(t) = C^{SL}(t)$ (if not, one can average over the two types of correlators or use the one with the smallest statistical fluctuations).
Thus in what follows we limit ourselves to consider three independent correlators: $C^{LL}(t)$, $C^{LS}(t)$ and $C^{SS}(t)$ sharing the same masses $M_i$ and, respectively, with real amplitudes $A_i^L A_i^L$, $A_i^L A_i^S$ and $A_i^S A_i^S$.

The ODE algorithm can be applied separately to each correlator $C^{f f^\prime}(t)$, namely Eq.~(\ref{eq:mass_eq_p}) becomes
\be
     \sum_{k=0}^{N-1} \overline{M}_{k^\prime k}^{f f^\prime} \overline{x}_k  = \overline{V}_{k^\prime}^{f f^\prime} ~ ,
     \label{eq:mass_eq_ffprime}
\ee
where the coefficients $\overline{x}_k $ should be independent of the specific choice of the correlator $C^{f f^\prime}(t)$.
In order to guarantee such an independence and to make a simultaneous use of the information contained in all the correlators we construct the variable $\widehat{\chi}_M^2$ given by
\be
    \widehat{\chi}_M^2 \equiv \sum_{f f^\prime=(LL,LS,SS)} ~ \sum_{n = N + 1}^{N_T - N} \frac{1}{[\sigma_n^{(0)}]^2} 
                                         \left\{ \sum_{k^\prime=0}^N \overline{x}_{k\prime} \left[ C_n^{(2k^\prime)} \right]^{f f^\prime} \right\}^2
    \label{eq:chi2_multiple}
\ee
and we impose the global constraints $\partial \widehat{\chi}_M^2 / \partial \overline{x}_k = 0$ for $k=0, 1, ... (N-1)$ with $\overline{x}_N = 1$.
It follows that the coefficients $\overline{x}_k$ should satisfy the linear system of equations
\be
     \sum_{k=0}^{N-1} \widehat{M}_{k^\prime k} \overline{x}_k  = \widehat{V}_{k^\prime} 
     \label{eq:mass_eq_multiple}
\ee
where
\bea
     \label{eq:mass_matrix_multiple}
     \widehat{M}_{k^\prime k} & \equiv & \overline{M}_{k^\prime k}^{LL} + \overline{M}_{k^\prime k}^{LS} + \overline{M}_{k^\prime k}^{SS} ~ , \\[2mm]
    \label{eq:vectorV_multiple}
    \widehat{V}_{k^\prime} & \equiv & \overline{V}_{k^\prime}^{LL} + \overline{V}_{k^\prime}^{LS} + \overline{V}_{k^\prime}^{SS}  ~ .
\eea

The solution of Eq.~(\ref{eq:mass_eq_multiple}) allows the determination of the roots of the polynomial (\ref{eq:PNz_p}) and consequently of the common masses $M_i$ .
Finally, the amplitudes $A_i^L$ and $A_i^S$ can be easily obtained after applying the procedure illustrated in Section \ref{sec:t-parity} to the individual correlators $C^{f f^\prime}(t)$.

\subsection{Filtering}
\label{sec:filtering}

Lattice correlation functions are defined in a (discretized) Euclidean space after performing the Wick rotation from the physical Minkowsky space.
After the rotation, however, correlation functions may contain exponential signals with masses lighter than the one relevant for the physical process under investigation.
A well-known example is the $K \to \pi \pi$ decay, where according to Ref.~\cite{Maiani:1990ca} the energy non-conserving matrix elements with the state of two pions at rest are involved.
In these cases, since at finite lattice volume the eigenvalues of the QCD Hamiltonian are discretized, a possible procedure is to try to subtract or, in other words, to filter out the {\it unwanted} signals.

The filtering of unwanted signals can be carried out after the determination of both masses and amplitudes for all the exponential signals present in the correlator.
In Appendix~\ref{sec:appendix} we describe a simple procedure, based on the ODE algorithm, that allows to filter out exponential signals from a given correlator without the need of determining the amplitudes of the unwanted signals.

\section{Inversion of the mass and amplitude matrices}
\label{sec:condition}

The numerical inversion of the mass and amplitude matrices can be carried out using standard methods, like the lower-upper decomposition (or factorization) method~\cite{LU}.
In what follows we will refer to the mass and amplitude matrices as defined in Section~\ref{sec:t-parity}.

If the interval of summation over $n$ in Eq.~(\ref{eq:mass_matrix_p}) is restricted to a single value of $n$, the matrix $M$ has a vanishing determinant and therefore it cannot be inverted.
Generally speaking, the sum over various values of $n$ protects against the singularity of the matrix $M$.
However, the mass matrix may still be close to singularity. 
As we shall see in this Section and in Section~\ref{sec:lattice}, the quasi-singularity represents on one hand side a positive feature, that allows the ODE algorithm to be sensitive to the fluctuations of the multiple exponential signals, and on the other hand side a limiting factor related to the presence of noise in the correlation function.

In the field of numerical analysis a condition number $\kappa(M)$ can be associated to the system of linear equations (\ref{eq:mass_eq_p}) and provides a bound on the relative error of the solution (\ref{eq:xk_p}) with respect to the relative error of the vector (\ref{eq:vectorV_p}).
The condition number $\kappa(M)$ is a property of the matrix $M$ and it is defined as~\cite{cond}
\be
    \kappa(M) \equiv || M^{-1} || \cdot || M || ~ ,
    \label{eq:condM}
\ee
where the symbol $|| M ||$ stands for the norm of the matrix $M$.
The latter can be defined in several ways and hereafter we adopt the Frobenius definition~\cite{LU}
\be
     || M || \equiv \sqrt{ \sum_{i=1}^N \sum_{j=1}^N |M_{ij}|^2} ~ .
     \label{eq:normM}
\ee
If the matrix $M$ is singular, then $\kappa(M) \to \infty$.
It is easy to show that
\be
    \frac{|| \delta x ||}{|| x ||} \leq \kappa(M) \frac{ || \delta V ||}{|| V ||} ~ ,
\ee
where $\delta x$ ($\delta V$) is the error on $x$ ($V$) and the vector norm is consistently defined as $ || x || = \sqrt{\sum_{i=1}^N x_i^2}$.
Thus, as a rule of thumb, when the condition number $\kappa(M)$ is equal to $\approx 10^d$, one may lose up to $d$ digits of accuracy in solving numerically Eq.~(\ref{eq:mass_eq_p}).

The condition number $\kappa(M)$ depends strongly on the dimension $N$ of the matrix $M$.
We have found that the appropriate handling of the numerical inversion of Eq.~(\ref{eq:mass_eq_p}) requires the use of multiple precision software.
In this work we have adopted the open-source software package MPFUN2015~\cite{mpfun}, which allows to change the precision level during run time.
We have used at least 32 digits of precision (quadruple precision) reaching 96 digits in the most severe cases.

Also the amplitude matrix (\ref{eq:amplitude_matrix_p}) needs to be inverted.
In this case the condition number $\kappa(A)$ turns out to be much smaller and it can be handled properly by adopting 32 digits of precision.

For finding the roots of the polynomial $P_N(z)$ we make use of the powerful, open-source software package MPSolve~\cite{MPSolve}.

\subsection{Fake data}
\label{sec:fake_data}

We now test the ODE algorithm by generating fake data for the correlator $C_n^{(0)}$ to be used as benchmarks.
We stress that, even if fake correlators represent an ideal situation different from the case of lattice correlators, it is nevertheless mandatory to check that our algorithm (as well as any other algorithm developed for fitting the temporal dependence of lattice correlators) is able to provide exact results in a controlled situation.

Let us start by considering a correlator with positive $t$-parity of the form 
\be
     C_n^{(0)} = \sum_{i=1}^N A_i \left[ e^{- a M_i n} + e^{- a M_i (N_T - n)} \right] ~ ,
     \label{eq:fake_Cn}
\ee
on a lattice with temporal extension $N_T = T / a = 96$ and spacing $a$ (the specific value of $a$ is not required since we work in lattice units). 
The fake data are generated by allowing the amplitudes $A_i$ and the masses $a M_i$ to fluctuate with uncertainties $\delta A_i$ and $\delta (aM_i)$, respectively, adopting (uncorrelated) Gaussian distributions to produce a total of 40 jackknives.
We remind that the ODE procedure is always applied to each single jackknife. 
The corresponding results provide central value and errors for the masses and the amplitudes according to the jackknife procedure.

The number of exponential signals in Eq.~(\ref{eq:fake_Cn}) is taken to be equal to $N = 12$.
The values chosen for the amplitudes and the masses are collected in Table~\ref{tab:bench1} together with their uncertainties $\delta A_i$ and $\delta (aM_i)$. 
The latter ones are taken to be equal to $1 \%$ of the corresponding amplitudes $A_i$ and masses $a M_i$.

\begin{figure}[htb!]
\begin{center}
\includegraphics[scale=0.415]{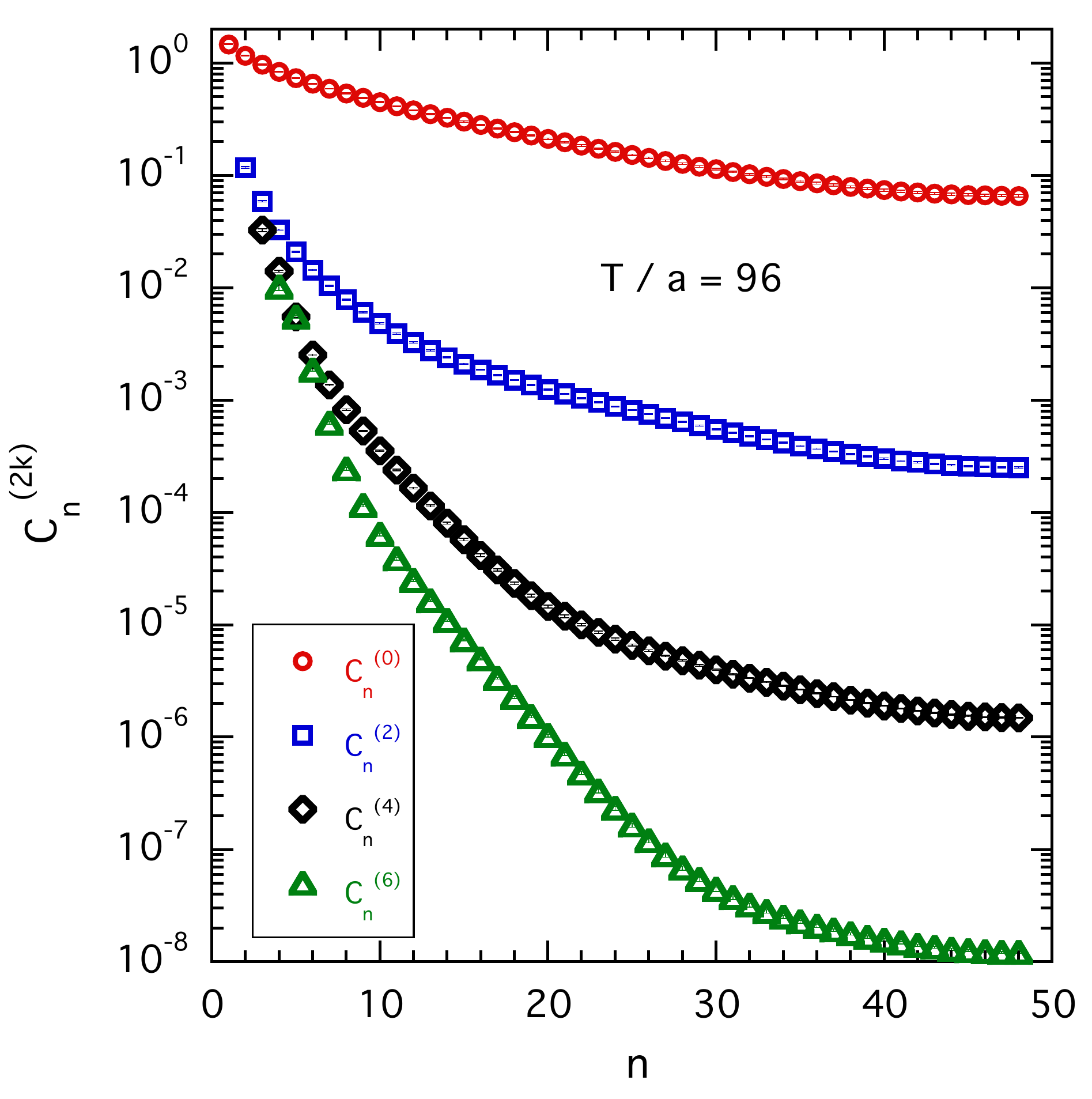}
\includegraphics[scale=0.415]{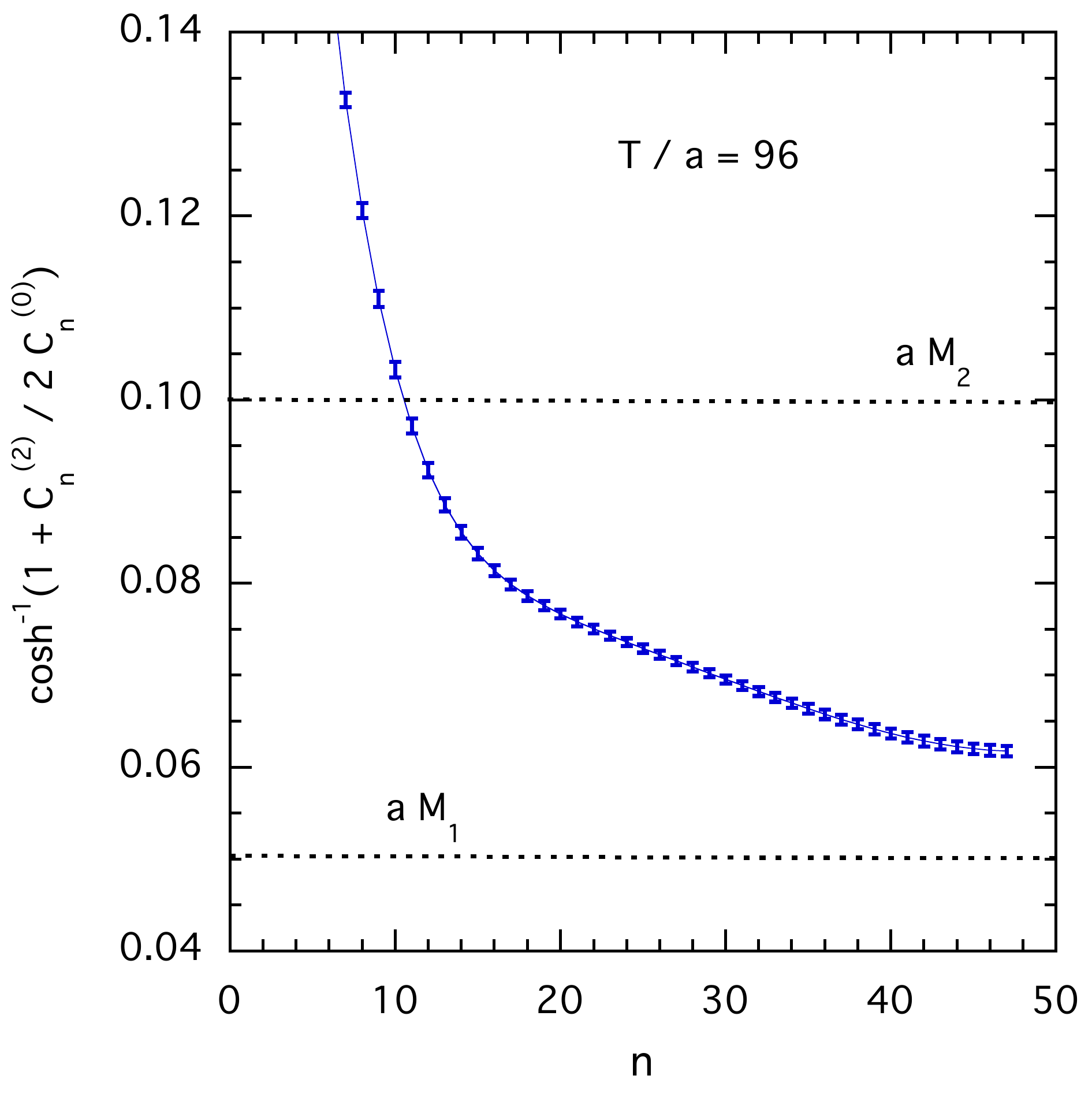}
\end{center}
\vspace{-0.50cm}
\caption{\it \small Left panel: time dependencies of the correlator $C_n^{(0)}$ and its derivatives $C_n^{(2k)}$ for $k = 1, 2, 3$ corresponding to the masses and amplitudes given in Table~\ref{tab:bench1}. Right panel: the effective mass $a M_{eff}^{(cosh)}$ (see Eq.~(\ref{eq:Meff_cosh})). The lower and upper dotted lines correspond to the locations of the ground and first excited states, respectively.}
\label{fig:bench1}
\end{figure}

The time dependencies of the correlator $C_n^{(0)}$, of the derivatives $C_n^{(2k)}$ for $k = 1, 2, 3$ and of the effective mass $a M_{eff}^{(cosh)}$ (see  Eq.~(\ref{eq:Meff_cosh})) are shown in Fig.~\ref{fig:bench1}.
The values of the lightest few masses have been chosen in such a way that at large time distances the ratio $C_n^{(2)} / C_n^{(0)}$ does not exhibit any plateaux within half of the temporal extension of the lattice. 
Thus, no direct information on the lightest mass can be extracted from the fake correlator using the standard effective mass methodology.

The numerical precision for generating the fake correlator (\ref{eq:fake_Cn}) with $N = 12$ is chosen to be 32 digits, i.e.~quadruple precision (see later on Table~\ref{tab:Nmax}), while all the {\it internal} ODE numerical calculations, namely the evaluation of the derivatives $C_n^{(2k)}$ for $k = 1, ...12$ as well as the numerical inversion of the mass and amplitude matrices of dimension $12 \times 12$, are performed using 64 digits (octuple precision).
The condition numbers for the mass and amplitude matrices (\ref{eq:mass_matrix_p}) and (\ref{eq:amplitude_matrix_p}) turn out to be equal to $\kappa(\overline{M}) \approx 10^{31}$ and $\kappa(\overline{A}) \sim 3 \cdot 10^5$, respectively.
Thus, with the octuple precision the numerical inversion of the mass and amplitude matrices is performed accurately.
\begin{table}[htb!]
\begin{center}
\begin{tabular}{||c|c|c||c|c|c||}
\hline
 $i$ & $a M_i$ & $A_i$ &  $i$ & $a M_i$ & $A_i$ \\
\hline \hline
 $~1$ & $0.05 \pm 0.0005$ & $~~0.30 \pm 0.0030$ &  $~7$ & $0.95 \pm 0.0095$ & $~~0.30 \pm 0.0030$ \\
\hline
 $~2$ & $0.01 \pm 0.0010$ & $~~0.70 \pm 0.0070$ &  $~8$ & $1.10 \pm 0.0110$ & $-0.45 \pm 0.0045$ \\
\hline
 $~3$ & $0.25 \pm 0.0025$ & $-0.10 \pm 0.0010$ &  $~9$ & $1.30 \pm 0.0130$ & $~~0.50 \pm 0.0050$ \\
\hline
 $~4$ & $0.40 \pm 0.0040$ & $~~0.80 \pm 0.0080$ &  $10$ & $1.55 \pm 0.0155$ & $~~0.30 \pm 0.0030$ \\
\hline
 $~5$ & $0.65 \pm 0.0065$ & $-0.30 \pm 0.0030$ &  $11$ & $1.80 \pm 0.0180$ & $~~0.20 \pm 0.0020$ \\
\hline
 $~6$ & $0.75 \pm 0.0075$ & $~~0.25 \pm 0.0025$ &  $12$ & $2.10 \pm 0.0210$ & $-0.55 \pm 0.0055$ \\
\hline 
\end{tabular}
\end{center}
\vspace{-0.25cm}
\caption{\it Values of the masses $a M_i$, amplitudes $A_i$ and their uncertainties $\delta (aM_i)$ and $\delta A_i$ adopted for generating the fake data for the correlator (\ref{eq:fake_Cn}) assuming $N = 12$. The temporal extension is $N_T = T / a = 96$.}
\label{tab:bench1}
\end{table}

We now apply the ODE algorithm [see Eqs.~(\ref{eq:mass_eq_p}-\ref{eq:vectorW_p}) of Section~\ref{sec:t-parity}] using the fake correlator $C_n^{(0)}$ and its derivatives $C_n^{(2k)}$ up to $k = N = 12$, i.e.~making the explicit use of the information on the number of exponential signals present in Eq.~(\ref{eq:fake_Cn}).
The case in which, besides the correlator $C_n^{(0)}$, the derivatives $C_n^{(2k)}$ up to $k = N_{ODE}$ with $N_{ODE} \neq N$ are considered, is postponed to Sections~\ref{sec:N_ODE} and~\ref{sec:N_ODE+}.
We remind that the definitions (\ref{eq:mass_matrix_p}-\ref{eq:vectorV_p}) and (\ref{eq:amplitude_matrix_p}-\ref{eq:vectorW_p}) correspond to the use of the values of the correlator $C_n^{(0)}$ in the full range $n = [1, \, N_T]$.

The ODE algorithm provides the values $X_i^{(ODE)} \equiv \{ a M_i^{(ODE)}, ~ A_i^{(ODE)} \}$ for the quantities $X_i \equiv \{ a M_i, ~ A_i \}$ and through the jackknife procedure the ODE errors $\delta X_i^{(ODE)} \equiv \{ \delta (a M_i^{(ODE)}), ~ \delta A_i^{(ODE)} \}$ for the corresponding uncertainties $\delta X_i \equiv \{ \delta (a M_i), ~ \delta A_i \}$ with $i = 1, ..., 12$.

In the case at hand, for the relative deviations we get
\bea
     \label{eq:bench}
     \left| \frac{X_i^{(ODE)} - X_i}{X_i} \right| \leq \Delta_{max} & \sim & 2 \cdot 10^{-13} ~ , \\[2mm]
     \label{eq:bench_delta}     
     \left| \frac{\delta X_i^{(ODE)} - \delta X_i}{\delta X_i} \right| \leq \delta \Delta_{max} & \sim & 2 \cdot 10^{-10} ~ .
\eea
We have also considered the cases in which the uncertainties $\delta (a M_i)$ and $\delta A_i$ are either increased up to $10 \%$ or decreased down to $0.1 \%$ of the corresponding masses $a M_i$ and amplitudes $A_i$.
In the former case the maximum relative deviations are $\Delta_{max} \sim 2 \cdot 10^{-13}$ and $\delta \Delta_{max} \sim 3 \cdot 10^{-11}$, while in the latter case we get $\Delta_{max} \sim 3 \cdot 10^{-13}$ and $\delta \Delta_{max} \sim 2 \cdot 10^{-9}$. 

The above findings illustrate very clearly that the ODE algorithm is able not only to determine precisely the central values of masses and amplitudes, but also to detect accurately the fluctuations generated in the fake correlator by the uncertainties of masses and amplitudes.
This ability is basically due to the huge value of the condition number $\kappa(\overline{M})$, i.e.~to the closeness of the mass matrix to singularity.
To deal with a huge value of $\kappa(\overline{M})$ is similar to the case of the study of a function $f(x)$ around a value $x = x_0$ where its derivative is huge.
Small variations of $x$ around $x_0$ can be detected, since they lead to large variations of the function $f(x)$.

As more exponential signals with masses above the heaviest one in Table~\ref{tab:bench1} are added in the fake correlator~(\ref{eq:fake_Cn}), both the condition numbers $\kappa(\overline{M})$ and $\kappa(\overline{A})$ as well as the maximum relative deviations $\Delta_{max}$ and $\delta \Delta_{max}$ increase quickly as shown in Table~\ref{tab:Deltamax}. 
\begin{table}[htb!]
\begin{center}
\begin{tabular}{||c||c|c||c|c||}
\hline
 $N$ & $\Delta_{max}$ &  $\delta \Delta_{max}$ & $\kappa(\overline{M})$ & $\kappa(\overline{A})$ \\
\hline \hline
$12$ & $\sim 2 \cdot 10^{-13}$ & $\sim 2 \cdot 10^{-10}$ & $\approx 10^{31}$ & $\sim 3 \cdot 10^5$ \\
\hline
 $14$ & $\sim 2 \cdot 10^{-11}$ & $\sim 4 \cdot 10^{-8}$\,\, & $\approx10^{38}$ & $\sim 5 \cdot 10^6$ \\
\hline
$16$ & $\sim 2 \cdot 10^{-4}$\,\, & $\sim 6 \cdot 10^{-2}$\,\, & $\approx10^{49}$ & $\sim 1 \cdot 10^9$ \\
\hline
\end{tabular}
\end{center}
\vspace{-0.25cm}
\caption{\it Values of the the maximum relative deviations $\Delta_{max}$ and $\delta \Delta_{max}$ versus the number $N$ of exponential signals included in the fake correlator $C_n^{(0)}$. The numerical precision of $C_n^{(0)}$ corresponds to $32$ digits (quadruple precision), while the internal numerical precision adopted in the ODE algorithm corresponds to $64$ digits (octuple precision). The temporal extension of the lattice is $N_T = T / a = 96$ and the uncertainties $\delta (a M_i)$ and $\delta A_i$ are equal to $1 \%$ of the corresponding masses and amplitudes $a M_i$ and $A_i$.}
\label{tab:Deltamax}
\end{table}

The quick rise of $\Delta_{max}$ and $\delta \Delta_{max}$ with $N$ is related to the impact of the numerical rounding of the fake correlator, while it does not depend on the numerical precision of the ODE method, which can be always kept at the desired level.
Indeed, the numerical accuracy of the fake correlator $C_n^{(0)}$ is {\it external} to the ODE method and it governs the maximum number of exponential signals that can be determined precisely (i.e.~within given values of the maximum relative deviations $\Delta_{max}$ and $\delta \Delta_{max}$).
In Table~\ref{tab:Nmax} we have collected the values of $N_{max}$ found indicatively for different levels of the numerical precision of the correlator $C_n^{(0)}$ while keeping $\Delta_{max} < 5 \cdot 10^{-9}$ and $\delta \Delta_{max} < 5 \cdot 10^{-6}$.
\begin{table}[htb!]
\begin{center}
\begin{tabular}{||c|c||c|c|c||}
\hline
 correlator precision & $N_{max}$ & $\kappa(\overline{M})$ & $\kappa(\overline{A})$ & internal precision \\
 (digits) & & & & (digits) \\
\hline \hline
$16$ & $6$ & $\approx 10^{14}$ & $\sim 3 \cdot 10^2$ & $32$  \\
\hline
 $32$ & $14$ & $\approx10^{38}$ & $\sim 5 \cdot 10^6$ & $64$\\
\hline
$64$ & $20$ & $\approx10^{74}$ & $\,\,\sim 7 \cdot 10^{14}$ & $96$ \\
\hline
\end{tabular}
\end{center}
\vspace{-0.25cm}
\caption{\it Values of the the maximum number $N_{max}$ of exponential signals that can be determined by the ODE algorithm keeping $\Delta_{max} < 5 \cdot 10^{-9}$ and $\delta \Delta_{max} < 5 \cdot 10^{-6}$ for different levels of the numerical precision of the correlator $C_n^{(0)}$. The third, fourth and fifth columns contain the condition numbers $\kappa(\overline{M})$, $\kappa(\overline{A})$ and the internal numerical precision adopted in the ODE algorithm, respectively. The temporal extension of the lattice is $N_T = T / a = 96$ and the uncertainties $\delta (a M_i)$ and $\delta A_i$ are equal to $1 \%$ of the corresponding masses and amplitudes $a M_i$ and $A_i$.}
\label{tab:Nmax}
\end{table}

We now want to discuss a particular set of masses and amplitudes for the correlator (\ref{eq:fake_Cn}), which is relevant for simulating a typical situation occurring when hadronic form factors are extracted from appropriate lattice correlators.
The usual procedure is to form a suitable ratio of correlation functions (typically, the ratio of 3-point and 2-point correlation functions), which at large time distances exhibits a plateau.
From the latter the hadronic form factor is obtained.
We stress that the above procedure requires that the temporal separation between the source and the sink should be large enough to allow the ground-state signal to be isolated. 
The contamination of excited states is usually investigated by varying the separation, which however may be computationally quite expensive. 

\begin{figure}[htb!]
\begin{center}
\includegraphics[scale=0.415]{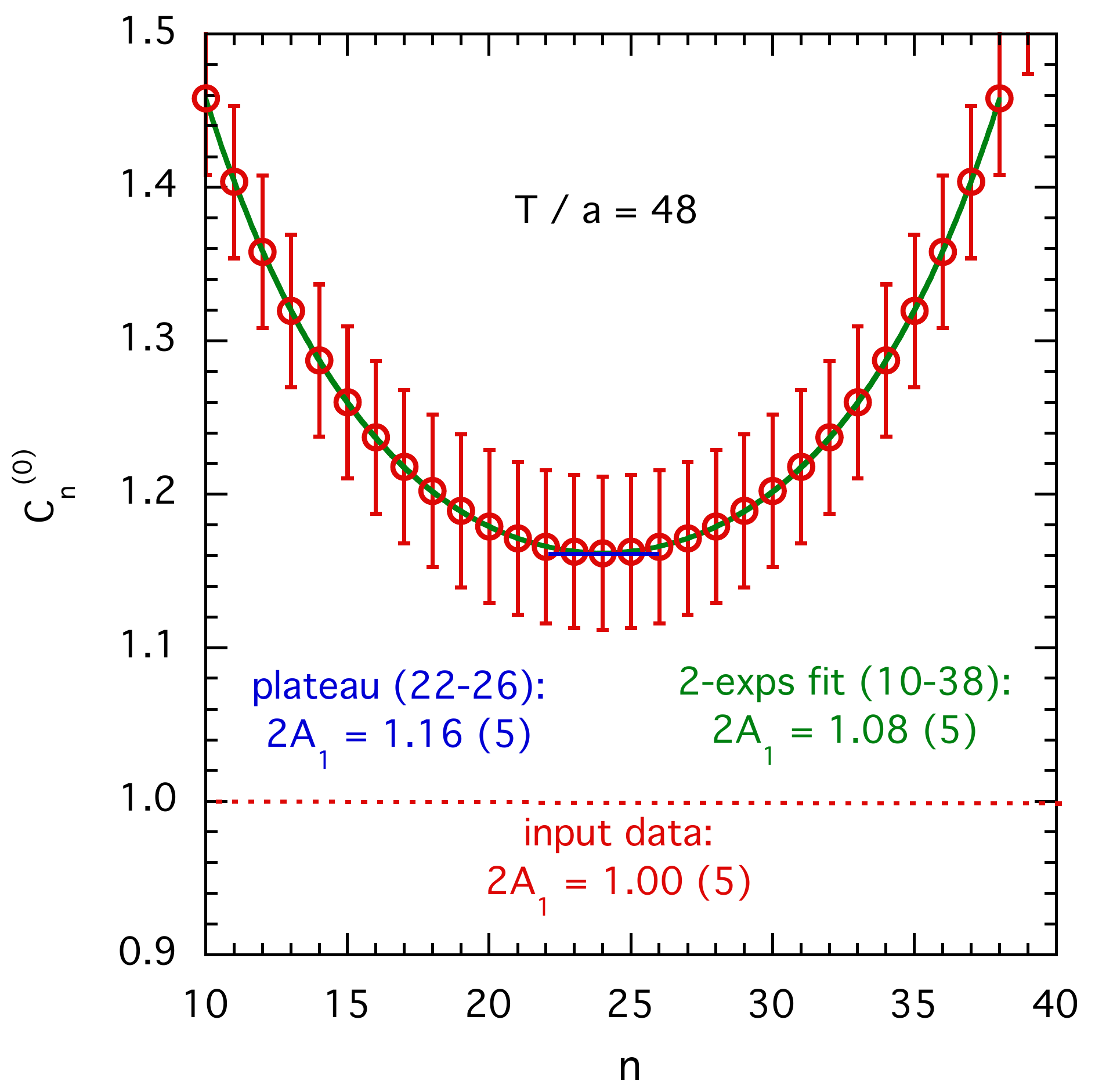}
\includegraphics[scale=0.415]{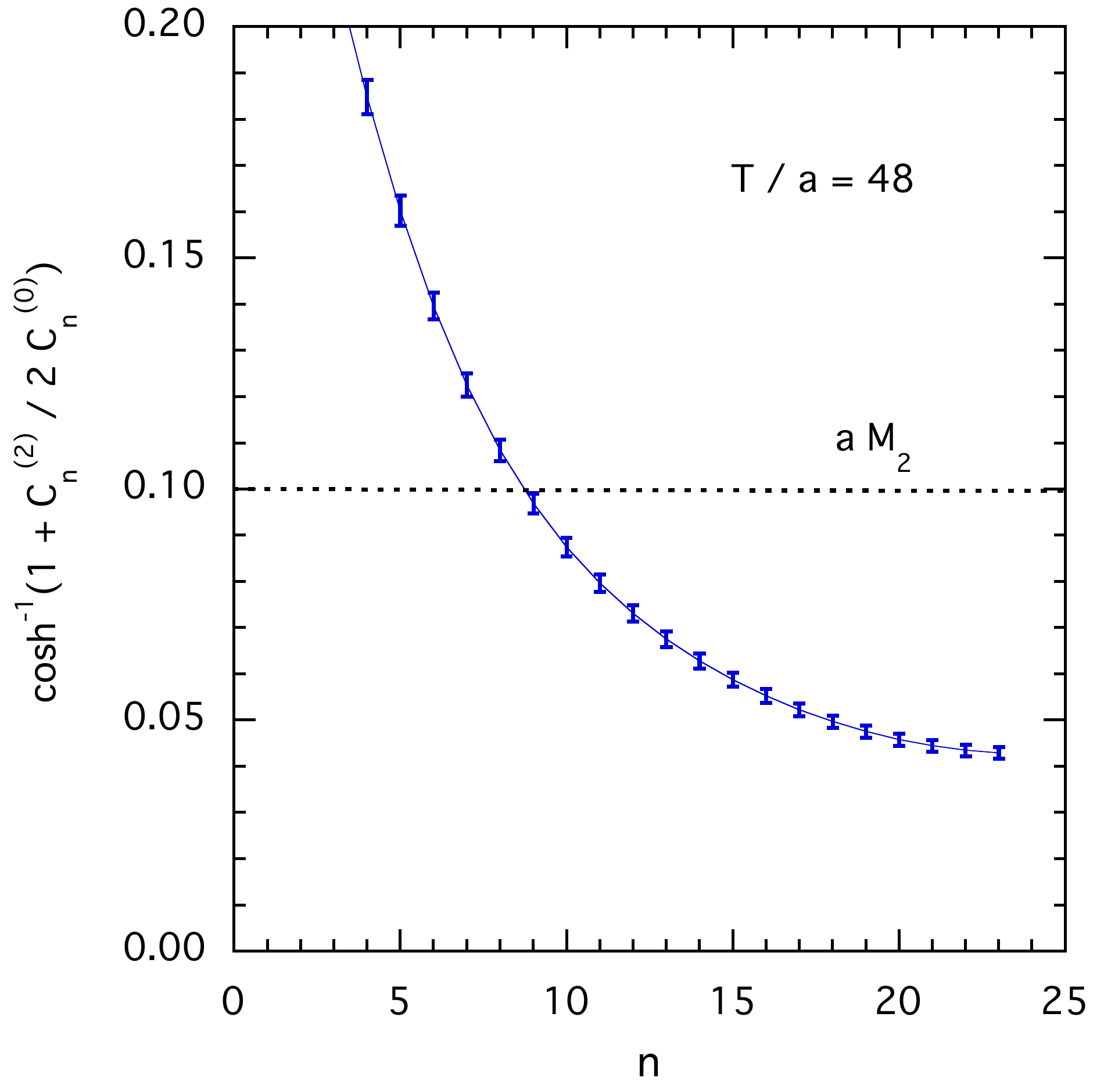}
\end{center}
\vspace{-0.50cm}
\caption{\it \small Left panel: time dependency of the correlator $C_n^{(0)}$ corresponding to the masses and amplitudes given in Table~\ref{tab:bench_3pts}. Right panel: the effective mass $a M_{eff}^{(cosh)}$ (see Eq.~(\ref{eq:Meff_cosh})). The dotted line corresponds to the location of the first excited state, $aM_2$, while the lightest state has a vanishing mass, $a M_1 = 0$.}
\label{fig:bench_3pts}
\end{figure}

In order to simulate the above situation we put the lightest mass in the fake correlator (\ref{eq:fake_Cn}) equal to zero, obtaining in this way a correlator that becomes constant at large time distances.
In Eq.~(\ref{eq:fake_Cn}) we consider $N = 4$ and a temporal extension equal to $N_T = T / a = 48$.
The values chosen for the amplitudes and the masses are collected in Table~\ref{tab:bench_3pts} together with their uncertainties $\delta A_i$ and $\delta (aM_i)$.
They have been chosen in such a way that at large time distances the correlator $C_n^{(0)}$ does exhibit a (short) plateau around half of the temporal extension of the lattice, as shown in the left panel of Fig.~\ref{fig:bench_3pts}.
\begin{table}[htb!]
\begin{center}
\begin{tabular}{||c|c|c||}
\hline
 $i$ & $a M_i$ & $A_i$ \\
\hline \hline
 $~1$ & $0.000 \pm 0.000$ & $0.500 \pm 0.025$ \\
\hline
 $~2$ & $0.100 \pm 0.001$ & $0.800 \pm 0.040$ \\
\hline
 $~3$ & $0.200 \pm 0.002$ & $1.000 \pm 0.050$ \\
\hline
 $~4$ & $0.500 \pm 0.005$ & $1.500 \pm 0.075$ \\
\hline 
\end{tabular}
\end{center}
\vspace{-0.25cm}
\caption{\it Values of the masses $a M_i$, amplitudes $A_i$ and their uncertainties $\delta (aM_i)$ and $\delta A_i$ adopted for generating the fake data for the correlator (\ref{eq:fake_Cn}) assuming $N = 4$. The temporal extension is $N_T = T / a = 48$.}
\label{tab:bench_3pts}
\end{table}
The plateau should in principle correspond to twice the amplitude $A_1$ of Table~\ref{tab:bench_3pts} (the factor of $2$ is due to the positive $t$-parity of the fake correlator), i.e.~$2 A_1 = 1.00 (5)$. 
Instead, the average of the correlator in the plateau range $n = [22, \, 26]$ provides the incorrect value $2 A_1 = 1.16 (5)$.
Moreover, by adopting a simple fit based on two exponential signals in the larger range $n = [10, \, 38]$ one gets an improved determination $2 A_1 = 1.08 (5)$, which however still differs from the input value. 
The temporal behavior of the effective mass $a M_{eff}^{(cosh)}$, given by Eq.~(\ref{eq:Meff_cosh}) and shown in the right panel of Fig.~\ref{fig:bench_3pts}, indicates clearly that the dominance of the ground-state signal, having a vanishing mass, is not yet reached. 
This is not at all a problem for the ODE algorithm, which is able to determine accurately all the masses and amplitudes of Table~\ref{tab:bench_3pts}. 

We point out that the above results illustrate an important, general feature of the ODE algorithm: it allows to extract accurately the ground-state signal without the need that the lattice temporal extension (or the temporal separation between the source and the sink for 3-point correlators) is large enough to allow the ground-state signal to be isolated.
In other words, the ODE method is able to remove properly the contamination of excited states also at relatively small values of the lattice temporal extension (or without changing the temporal separation between the source and the sink for 3-point correlators).

Let us now consider the case of a correlator containing poles with arbitrary multiplicity, namely
\be
     C_n^{(0)} = \sum_{i=1}^{\overline{N}} \sum_{\mu=0}^{\mu_i-1} A_{i \mu} \, a^\mu \, \left[ n^\mu e^{-aM_i n} + 
                        (N_T - n)^\mu e^{-aM_i (N_T - n)} \right] ~ ,
     \label{eq:fake_C0n_mu}
\ee
where $\mu_i$ is the multiplicity of the $i$-th exponential with $i = 1, 2, ... \overline{N}$ and $\sum_{i=1}^{\overline{N}} \mu_i = N$.
The values chosen for the masses $aM_i $, the amplitudes $A_{i \mu}$ and the multiplicities $\mu_i$ are collected in Table~\ref{tab:bench2} together with the corresponding uncertainties $\delta (aM_i)$ and $\delta A_{i \mu}$.
There are four single poles, two double poles and a quadruple pole for a total of $N = 12$ exponential signals.
The relative uncertainties are taken to be equal to $1 \%$ in the case of the masses and $5 \%$ for the amplitudes.
The quadruple precision (32 digits) is used for evaluating the fake correlator (\ref{eq:fake_C0n_mu}) and the octuple precision (64 digits) for the internal ODE calculations.
\begin{table}[htb!]
\begin{center}
{\small
\begin{tabular}{||c|c|c|c||c|c|c|c||}
\hline
 $i$ & $a M_i$ & $\mu_i$ & $A_{i \mu}$ &  $i$ & $a M_i$ & $\mu_i$ & $A_{i \mu}$ \\
\hline \hline
 $1$ & $\,\,0.15 \pm 0.0015$ & $2$ &  $~\,\,0.050 \pm 0.0025$ & $5$ & $1.20 \pm 0.012$ & $4$ & $~\,\,0.50 \pm 0.025$ \\
\cline{4-4} \cline{8-8}
        &                                    &        & $~\,\,0.002 \pm 0.0010$ &         &                              &        & $~\,\,0.60 \pm 0.030$ \\
\cline{1-4} \cline{8-8}
 $2$ & $0.30 \pm 0.003$       & $1$ & $-0.10 \pm 0.005$ &                  &                              &        & $-0.20 \pm 0.010$ \\
\cline{1-4} \cline{8-8}
 $3$ & $0.60 \pm 0.006$       & $2$ & $-0.20 \pm 0.010$ &                  &                              &       & $~\,\,1.00 \pm 0.050$ \\
\cline{4-8}
        &                                    &        & $\,\,0.500 \pm 0.025$ & $6$ & $1.50 \pm 0.015$ & $1$ & $\,\,1.50 \pm 0.075$ \\
\cline{1-8} 
 $4$ & $0.90 \pm 0.009$       & $1$ & $\,\,1.000 \pm 0.050$ & $7$ & $2.00 \pm 0.020$ & $1$ & $\,\,2.00 \pm 0.100$ \\
\hline 
\end{tabular}
}
\end{center}
\vspace{-0.25cm}
\caption{\it Values of the masses $a M_i$, amplitudes $A_{i \mu}$, multiplicities $\mu_i$ and the uncertainties $\delta (aM_i)$ and $\delta A_{i \mu}$ adopted for generating the fake data for the correlator (\ref{eq:fake_C0n_mu}). The temporal extension is $N_T = T / a = 96$.}
\label{tab:bench2}
\end{table}

\begin{figure}[htb!]
\begin{center}
\includegraphics[scale=0.415]{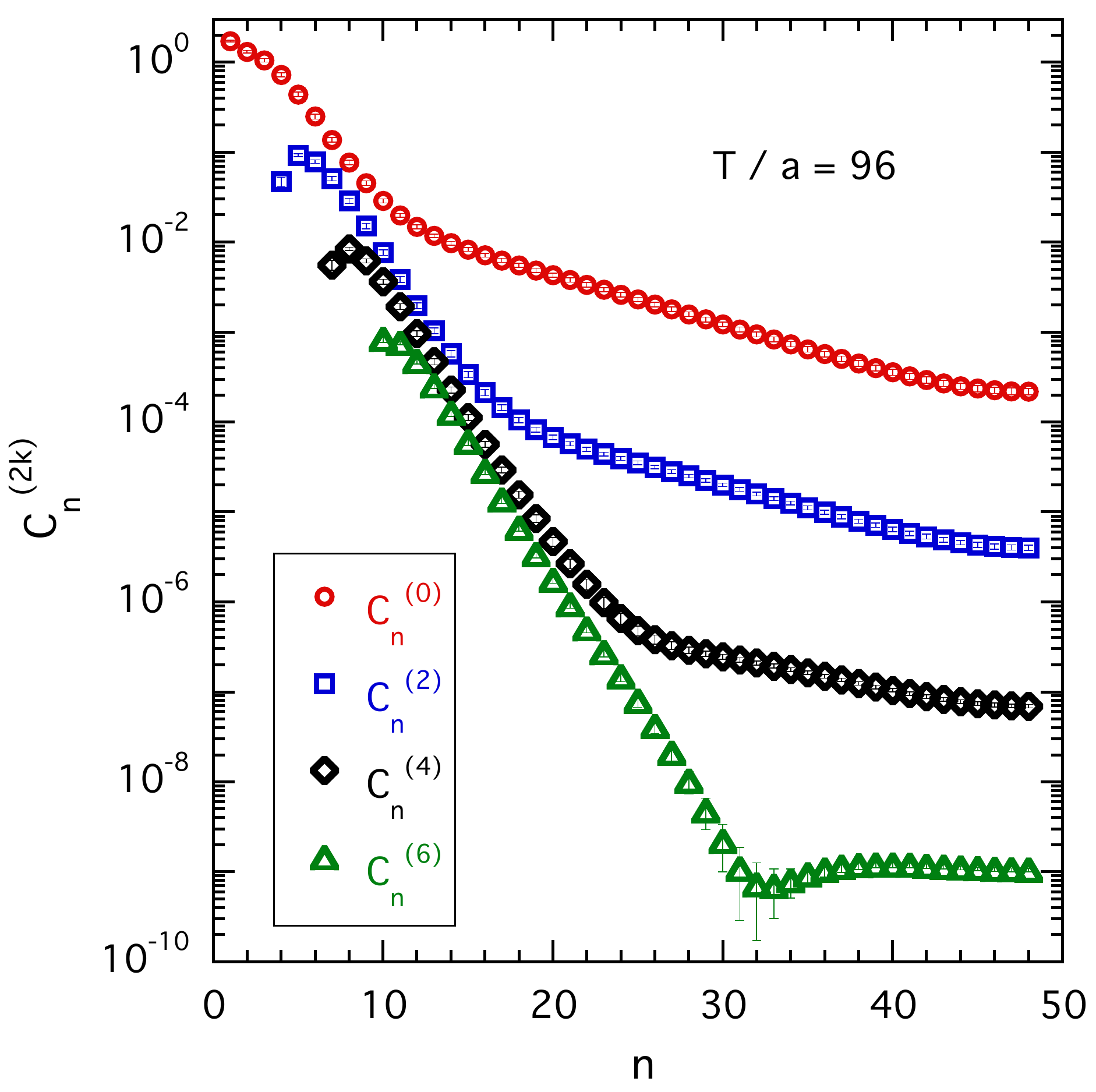}
\includegraphics[scale=0.415]{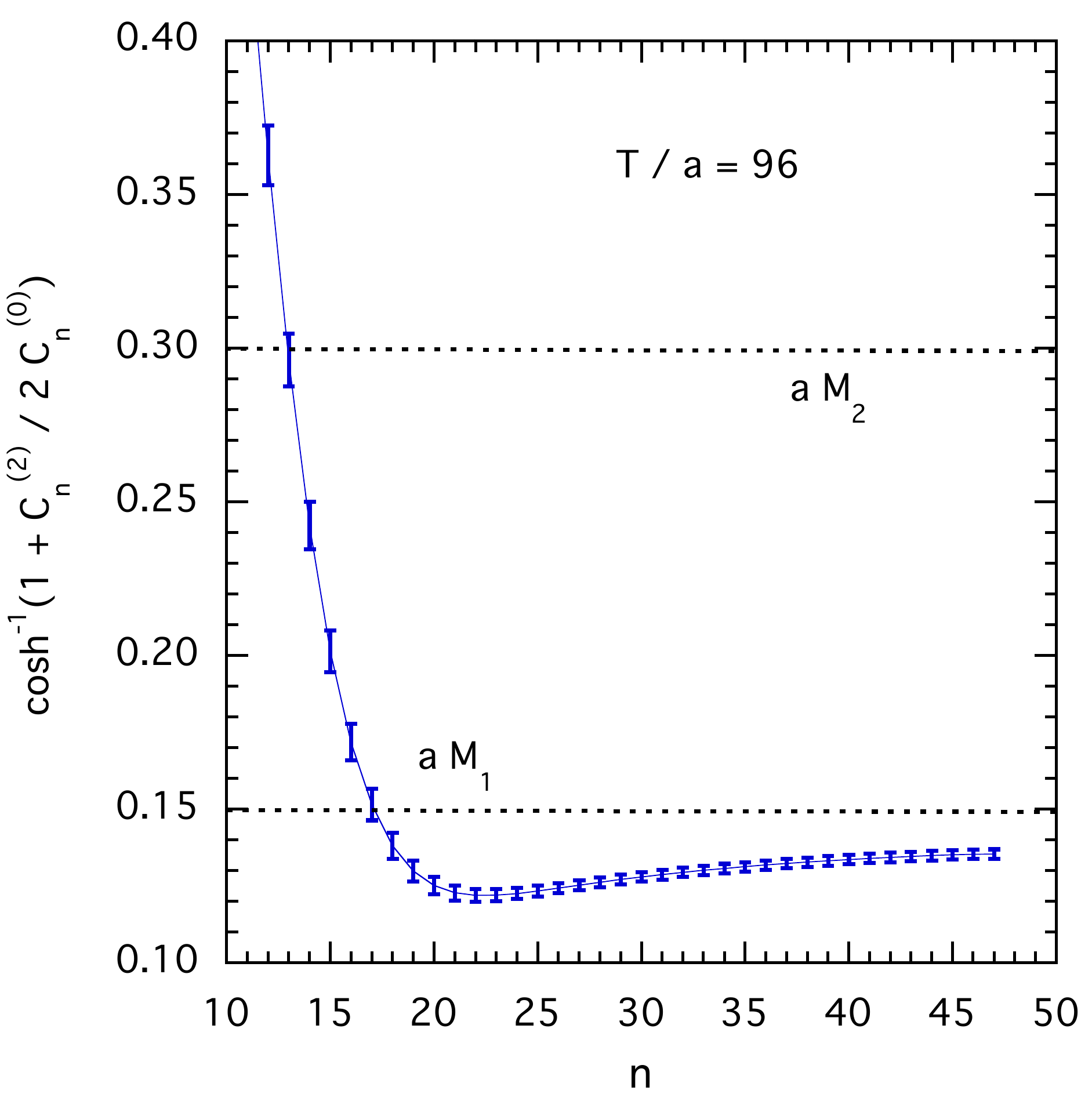}
\end{center}
\vspace{-0.50cm}
\caption{\it \small As in Fig.~\ref{fig:bench1} but in the case of the fake correlator (\ref{eq:fake_C0n_mu}) with the masses and amplitudes given in Table~\ref{tab:bench2}.}
\label{fig:bench2}
\end{figure}

The time dependencies of $C_n^{(2k)}$ for $k = 0, ... 3$ and the one of the effective mass $a M_{eff}^{(cosh)}$ are shown in Fig.~\ref{fig:bench2}.
It can be seen that at large time distances the effective mass~(\ref{eq:Meff_cosh}) does not exhibit a plateau and, therefore, the lightest mass cannot be determined accurately by the standard effective mass procedure.

We now apply the ODE algorithm using all the derivatives $C_n^{(2k)}$ up to $k = N = 12$.
The condition numbers are found to be $\kappa(\overline{M}) \approx 10^{28}$ and $\kappa(\overline{A}) \sim 2 \cdot 10^{12}$.
Multiple roots are now present in the polynomial (\ref{eq:PNz_mu}) due to the multiple poles in the fake correlator.
The ODE algorithm determines accurately all the masses, amplitudes and multiplicities of Table~\ref{tab:bench2}.
For the maximum relative deviations [see Eqs.~(\ref{eq:bench}-\ref{eq:bench_delta})] we get $\Delta_{max} \sim 10^{-12}$ and $\delta \Delta_{max} \sim 2 \cdot 10^{-10}$, which means that all the masses and amplitudes of Table~\ref{tab:bench2} as well as the multiple pole structure of the correlator (\ref{eq:fake_C0n_mu}) are determined very precisely.

An interesting case is represented by the following fake correlator
\be
     C_n^{(0)} = \sum_{m=1}^6 \left\{ \left[ \widetilde{A}_m+ a \widetilde{B}_m n \right] e^{- a \widetilde{M}_m n} + 
                        \left[ \widetilde{A}_m + a \widetilde{B}_m (N_T - n) \right] e^{- a \widetilde{M}_m (N_T - n)} \right\} 
     \label{eq:fake_C0n_stg}
\ee
composed by six double poles with masses and amplitudes given in Table~\ref{tab:bench3}.
Three signals ($m = 1, 3, 5$) are non-oscillating double poles (i.e.~the corresponding masses are real), while the other three signals ($m = 2, 4, 6$) are oscillating double poles (i.e.~the corresponding masses contain an imaginary part equal to $-\pi$).
The numerical precision for generating the fake correlator (\ref{eq:fake_C0n_stg}) is again 32 digits (quadruple precision).
\begin{table}[htb!]
\begin{center}
\begin{tabular}{||c|c|c|c||}
\hline
 $m$ & $a \widetilde{M}_m$ & $\widetilde{A}_m$ & $\widetilde{B}_m$ \\
\hline \hline
 $1$ & $0.05 \pm 0.0020$ & $~~1.00 \pm 0.0020$ & $~~0.50 \pm 0.0040$ \\
\hline
 $2$ & $0.20 \pm 0.0030 - i\pi$ & $~~0.70 \pm 0.0030$ & $-0.20 \pm 0.0060$ \\
\hline
 $3$ & $0.55 \pm 0.0020$ & $~~0.80 \pm 0.0030$ & $-0.30 \pm 0.0060$ \\
\hline
 $4$ & $0.80 \pm 0.0040 - i\pi$ & $~~0.25 \pm 0.0080$ & $~~0.30 \pm 0.0060$ \\
\hline
 $5$ & $1.20 \pm 0.0050$ & $-0.45 \pm 0.0100$ & $~~0.50 \pm 0.0120$ \\
\hline
 $6$ & $1.60 \pm 0.0020 - i\pi$ & $~~0.30 \pm 0.0140$ & $~~0.30 \pm 0.0140$ \\
\hline
\end{tabular}
\end{center}
\vspace{-0.25cm}
\caption{\it Values of the masses $a \widetilde{M}_m$ and the amplitudes $\widetilde{A}_m$ and $\widetilde{B}_m$ ($m = 1, ..., 6$) together with their uncertainties adopted for generating the fake data for the correlator (\ref{eq:fake_C0n_stg}). The temporal extension is $N_T = T / a = 96$.}
\label{tab:bench3}
\end{table}

The application of the ODE algorithm (with 64 digits of precision) to the fake correlator $C_n^{(0)}$ and its derivatives $C_n^{(2k)}$ with $k = 1, ..., 12$ is successful. 
All the six double poles are correctly found: three roots $\widetilde{z}_m \equiv 2 \left[ \mbox{cosh}(a \widetilde{M}_m) - 1 \right]$ are positive (corresponding to real masses) and the other three roots are less than $-4$ (corresponding to masses with an imaginary part equal to $-\pi$).
The condition numbers of the mass and amplitude matrices are $\sim 5 \cdot 10^{30}$ and $\sim 3 \cdot 10^{10}$, respectively.
The accuracy of the ODE results for all the masses, amplitudes and their uncertainties is better than $\sim 1$ ppb.

Results with the same quality can be obtained in the case of multiple correlators (see Section~\ref{sec:LS}), which we do not report here for sake of brevity.
We just mention that a well-established procedure to deal with multiple correlators is represented by the method based on the Generalized Eigenvalue Problem (GEVP)~\cite{Blossier:2009kd}.
With this method it is possible to extract masses and amplitudes of both ground and excited states.
In particular, with four correlators, generated using two different interpolating fields, the ground and the first excited states can be determined.
As the number of excited states increases, the number of interpolating fields and correspondingly the number of correlators should be increased.
This is at variance with the ODE method, which is able to detect properly many exponential signals independently on the number of correlators used.

\subsection{ODE analyses with $N_{ODE} < N$}
\label{sec:N_ODE}

In this Section we address the case in which the total number of derivatives $C_n^{(2k)}$ included in the construction of the mass and amplitude matrices is less than the total number of exponential signals included in the fake data for the correlator $C_n^{(0)}$.
For the latter we use hereafter Eq.~(\ref{eq:fake_Cn}) with masses and amplitudes given in Table~\ref{tab:bench1} for $N = 12$.

We apply the ODE algorithm using $C_n^{(0)}$ and its derivatives $C_n^{(2k)}$ for $k = 1, ..., N_{ODE}$ with $N_{ODE} < N$, so that the mass and amplitude matrices have now dimension $N_{ODE} \times N_{ODE}$ . 
Furthermore, we consider that the range of analysis may not include all the values of the time distance at which the correlator $C_n^{(0)}$ is known, i.e.~we may use a range $n = [n_0 + 1, \, N_T - n_0]$, where the integer $n_0$ may be larger than $0$, at variance with what done in the previous Sections.

We write down explicitly the main ingredients of the ODE algorithm to take into account the use of a limited range of analysis $[n_0 + 1, \, N_T - n_0]$.
The mass matrix (\ref{eq:mass_matrix_p}) is replaced by
\be
      \overline{M}_{k^\prime k} \equiv \sum_{n = N_{ODE} + n_0 + 1}^{N_T - N_{ODE} - n_0} ~ \frac{C_n^{(2k^\prime)} \, C_n^{(2k)}}{[\sigma_n^{(0)}]^2}
      \label{eq:mass_matrix_ODE}
\ee
with $k, k^\prime = 0, 1, ..., (N_{ODE}-1)$ and the vector (\ref{eq:vectorV_p}) is now given by
\be
    \overline{V}_{k^\prime} \equiv - \sum_{n = N_{ODE} + n_0 + 1}^{N_T - N_{ODE} - n_0} ~ \frac{C_n^{(2k^\prime)} \, C_n^{(2N_{ODE})}}{[\sigma_n^{(0)}]^2} ~ .
    \label{eq:vectorV_ODE}
\ee
The solution of the ODE conditions
\be
     \sum_{k=0}^{N_{ODE}-1} \overline{M}_{k^\prime k} \overline{x}_k  = \overline{V}_{k^\prime} 
     \label{eq:xK_ODE}
\ee
provides the coefficients $\overline{x}_k$, which are used to construct the polynomial of degree $N_{ODE}$
\be
    \overline{P}_{N_{ODE}}(z) =  \sum_{k=0}^{N_{ODE}-1} \overline{x}_k z^k + z^N = \prod_{j=1}^{N_{ODE}} (z - \overline{z}_j^{ODE}) 
     \label{eq:PNz_ODE}
\ee
having its roots at $z = \overline{z}_j^{ODE} \equiv 2 \left[ \mbox{cosh}(a M_j^{ODE}) - 1 \right]$.
Once the $N_{ODE}$ masses $a M_j^{ODE}$ have been determined, the corresponding amplitudes $A_j^{ODE}$ can be obtained by solving the linear system of equations
\be
    \sum_{j^\prime=1}^{N_{ODE}} \overline{A}_{j j^\prime} A_{j^\prime}^{ODE} = \overline{W}_j ~ ,
     \label{eq:amplitude_eq_ODE} 
\ee
where
\bea
     \label{eq:amplitude_matrix_ODE}
     \overline{A}_{j j^\prime} & \equiv & \sum_{k=0}^{N_{ODE}} \sum_{n=k+1+n_0}^{N_T-k-n_0} (\overline{z}_j^{ODE} ~ \overline{z}_{j^\prime}^{ODE})^k \, 
                                                            \frac{\overline{f}_n^{(j)} \, \overline{f}_n^{(j^\prime)}}{[\sigma_n^{(2k)}]^2} ~ , \\[2mm]
     \label{eq:vectorW_ODE}
     \overline{W}_j & \equiv & \sum_{k=0}^{N_{ODE}} \sum_{n=k+1+n_0}^{N_T-k-n_0} (\overline{z}_j^{ODE})^k \, 
                                              \frac{\overline{f}_n^{(j)} \, C_n^{(2k)}}{[\sigma_n^{(2k)}]^2} ~ 
\eea
with\footnote{Eq.~(\ref{eq:fn_j_ODE}) holds for signals with real mass and multiplicity equal to $1$, as assumed in the fake data (\ref{eq:fake_Cn}). Later on we generalize Eq.~(\ref{eq:fn_j_ODE}) to the case of  imaginary ODE masses [see Eqs.~(\ref{eq:fn_j_ODE_Im}-\ref{eq:sn_j})] and multiple poles [see Eqs.~(\ref{eq:Cn_ODE_mu}-\ref{eq:fn_j_Im_mu})].}
\be
    \overline{f}_n^{(j)} \equiv e^{- aM_j^{ODE} n} + e^{- aM_j^{ODE} (N_T - n)} ~ .
    \label{eq:fn_j_ODE}
\ee

Generally speaking we expect that the ODE masses $a M_j^{ODE}$ represent an approximation of the lighter $N_{ODE}$ masses $a M_j$ included in the correlator $C_n^{(0)}$.
The signals with masses $a M_i$ above $a M_{N_{ODE}}$ (i.e.~for $i = [ N_{ODE} +1, \, N]$) certainly affect the elements of both the mass matrix $ \overline{M}$ and the vector $\overline{V}$. 
A kind of interaction among the roots may be generated, which moves the location of the ODE roots $\overline{z}_j^{ODE}$ of the polynomial (\ref{eq:PNz_ODE}) away from the exact ones $\overline{z}_j = 2 \left[ \mbox{cosh}(a M_j) - 1 \right]$, likely to higher values.
The effect of the above interaction depends not only on the signals having the higher masses, but also on the range of the analysis, i.e.~on the value of $n_0$. 
The difference between $a M_j^{ODE}$ and $a M_j$ is expected to decrease as $n_0$ increases, since the impact of the signals with higher masses decreases as the time distance increases.

This is indeed the phenomenology we observe.
In Tables~\ref{tab:ODE8_1} and~\ref{tab:ODE8_2} we have collected the results obtained by applying the ODE algorithm assuming $N_{ODE} = 8$ in the ranges $n = [5,  \, 92]$ ($n_0 = 4$) and $n = [10,  \, 87]$ ($n_0 = 9$), respectively.
\begin{table}[htb!]
\begin{center}
\begin{tabular}{||c|c|c||}
\hline
 $j$ & $a M_j^{ODE}$ & $A_j^{ODE}$ \\
\hline \hline
 $1$ & $0.04995 \pm 0.00050$ & $~~0.29883 \pm 0.00307$ \\
\hline
 $2$ & $0.09986 \pm 0.00100$ & $~~0.70001 \pm 0.00697$ \\
\hline
 $3$ & $0.25781 \pm 0.00323$ & $-0.11321 \pm 0.00421$ \\
\hline
 $4$ & $0.39552 \pm 0.00418$ & $~~0.79056 \pm 0.00911$ \\
\hline 
$5$ & \small{$0.72065 \pm 0.00522 + i \, (0.110 \pm 0.025)$} & $~~0.05472 \pm 0.01392$ \\
\hline
$6$ & \small{$0.72065 \pm 0.00522 - i \, (0.110 \pm 0.025)$} & $-0.02925 \pm 0.00472$\\
\hline
$7$ & $1.5015 \pm 0.01899$ & $~~0.67175 \pm 0.03405$ \\
\hline
$8$ & $2.1090 \pm 0.05427$ & $-0.43229 \pm 0.01820$ \\
\hline
\end{tabular}
\end{center}
\vspace{-0.25cm}
\caption{\it Values of the masses $a M_j^{ODE}$, amplitudes $A_j^{ODE}$ and their uncertainties obtained by applying the ODE algorithm with $N_{ODE} = 8$ to the correlator (\ref{eq:fake_Cn}) containing $N = 12$ exponential signals with masses and amplitudes given in Table~\ref{tab:bench1}. The range of the ODE analysis is $n = [n_0 + 1, \, N_T - n_0] = [5, \, 92]$. All the roots of the polynomial (\ref{eq:PNz_ODE}) are found to have multiplicity equal to $\mu_i = 1$}
\label{tab:ODE8_1}
\end{table}
\begin{table}[htb!]
\begin{center}
\begin{tabular}{||c|c|c||c|c||}
\hline
 $j$ & $a M_j^{ODE}$ & $A_j^{ODE}$ & $a M_j$ & $A_j$ \\
\hline \hline
 $1$ & $0.05000 \pm 0.00050$ & $~~0.30000 \pm 0.00300$ & $0.05 \pm 0.0005$ & $~~0.30 \pm 0.0030$ \\
\hline
 $2$ & $0.10000 \pm 0.00100$ & $~~0.70000 \pm 0.00700$ & $0.01 \pm 0.0010$ & $~~0.70 \pm 0.0070$ \\
\hline
 $3$ & $0.24993 \pm 0.00250$ & $-0.09987 \pm 0.00100$ & $0.25 \pm 0.0025$ & $-0.10 \pm 0.0010$ \\
\hline
 $4$ & $0.40009 \pm 0.00401$ & $~~0.80080 \pm 0.00800$ & $0.40 \pm 0.0040$ & $~~0.80 \pm 0.0080$ \\
\hline 
$5$ & $0.63611 \pm 0.00575$ & $-0.22323 \pm 0.01413$ & $0.65 \pm 0.0065$ & $-0.30 \pm 0.0030$ \\
\hline
$6$ & $0.82329 \pm 0.00451$ & $~~0.30990 \pm 0.02311$ & $0.75 \pm 0.0075$ & $~~0.25 \pm 0.0025$ \\
\hline
$7$ & $1.17580 \pm 0.05828$ & $-0.21524 \pm 0.08739$ & $0.95 \pm 0.0095$ & $~~0.30 \pm 0.0030$ \\
\hline
$8$ & $1.40830 \pm 0.01651$ & $~~0.72549 \pm 0.08468$ & $1.10 \pm 0.0110$ & $-0.45 \pm 0.0045$ \\
\hline
\end{tabular}
\end{center}
\vspace{-0.25cm}
\caption{\it The same as in Table~\ref{tab:ODE8_1}, but for a range of the ODE analysis equal to $n = [n_0 + 1, \, N_T - n_0] = [10, \, 87]$. In the fourth and fifth columns the values of the masses and amplitudes for the first eight signals of Table~\ref{tab:bench1} are shown for ease of comparison.}
\label{tab:ODE8_2}
\end{table}
First of all it can be seen that in Table 5 there are masses with a non-vanishing imaginary part (namely $j = 5$ and $j = 6$).
They are complex conjugate, since the coefficients of the polynomial (\ref{eq:PNz_ODE}) are real.
Their presence is a signature that in the fake data of the correlator $C_n^{(0)}$ there are more than eight exponential signals.
In order to calculate the amplitudes $A_j^{ODE}$ we have to take into account the presence of conjugate masses, which appear in pairs: one with a positive value of the imaginary part and the other with an opposite (negative) value of the imaginary part. 
Thus, we replace Eq.~(\ref{eq:fn_j_ODE}) with
\be
    \overline{f}_n^{(j)} \equiv s_n^j e^{- a {\rm Re}(M_j^{ODE}) n} + s_{N_T -n}^j e^{- a {\rm Re}(M_j^{ODE}) (N_T - n)} ~ .
    \label{eq:fn_j_ODE_Im}
\ee
where\footnote{Eq.~(\ref{eq:fn_j_ODE_Im}) holds for signals with multiplicity equal to $1$. Later on, Eq.~(\ref{eq:fn_j_ODE_Im}) will be generalized to the case of multiple poles [see Eqs.~(\ref{eq:Cn_ODE_mu}-\ref{eq:fn_j_Im_mu})].}
\bea
    s_n^j & \equiv & \mbox{cos}[a{\rm Im(M_j^{ODE})} n] \qquad \qquad \mbox{for } {\rm Im(M_j^{ODE}) > 0} ~ , \nonumber \\[2mm]
             & \equiv & \frac{\mbox{sin}[a{\rm Im(M_j^{ODE})} n]}{a{\rm Im(M_j^{ODE})}} \qquad \qquad \mbox{for } {\rm Im(M_j^{ODE}) < 0} ~ .
    \label{eq:sn_j}
\eea
In the limit ${\rm Im(M_j^{ODE}) \to 0}$ the two exponential signals become a double pole and, indeed, Eq.~(\ref{eq:sn_j}) provides the appropriate limits $s_n^j \to 1$ for ${\rm Im(M_j^{ODE}) \to 0^+}$ and $s_n^j \to n$ for ${\rm Im(M_j^{ODE}) \to 0^-}$.

The lighter four exponential signals are reproduced within the uncertainties when the range of the analysis is equal to $n = [5, \, 92]$, and quite precisely already when $n = [10, \, 87]$.
On the contrary the heavier four signals are not reproduced at all for $n = [5, \, 92]$ and only the mass of the fifth signal matches the exact one within the uncertainty for $n = [10, \, 87]$.

Though not all the ODE masses and amplitudes reproduce the input values, the fake data of the correlator $C_n^{(0)}$ are reasonably reproduced by its ODE representation
\be
     C_n^{(ODE)} = \sum_{j=1}^{N_{ODE}} A_j^{ODE} ~ \overline{f}_n^{(j)} ~ .
     \label{eq:Cn_ODE}
\ee
We find that the absolute value of the residue $[C_n^{(0)} - C_n^{(ODE)}]$ does not exceed $2 \cdot 10^{-3} \, \sigma_n^{(0)}$ and $10^{-5} \, \sigma_n^{(0)}$ for the analyses in the ranges $n = [5, \, 92]$ and $n = [10, \, 87]$, respectively.
This means that small values of the residue do not guarantee that masses and amplitudes are properly reproduced.

\begin{figure}[htb!]
\begin{center}
\includegraphics[scale=0.85]{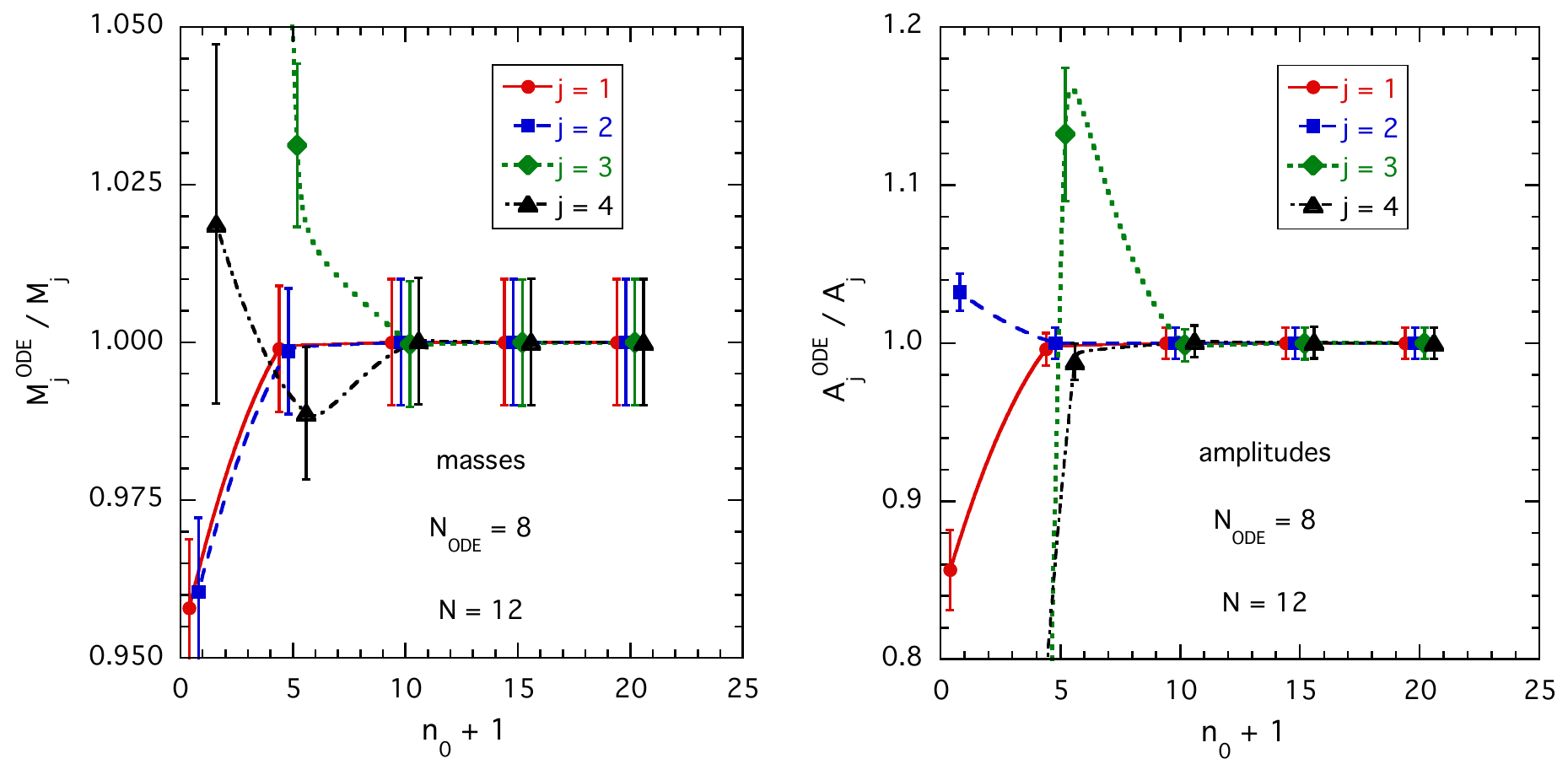}
\end{center}
\vspace{-0.75cm}
\caption{\it \small Ratios $M_j^{ODE} / M_j$ (left panel) and $A_j^{ODE} / A_j$ (right panel) for the lighter four states, obtained by the ODE algorithm assuming $N_{ODE} = 8$ for the analysis of the correlator (\ref{eq:fake_Cn}) containing $N = 12$ exponential signals (see Table~\ref{tab:bench1}), versus the value of $n_0 + 1$, which defines the range of the analysis $n = [n_0 + 1, \, N_T - n_0]$. The results for the various states are slightly shifted horizontally for better readability.}
\label{fig:ODE8_1}
\end{figure}

\begin{figure}[htb!]
\begin{center}
\includegraphics[scale=0.85]{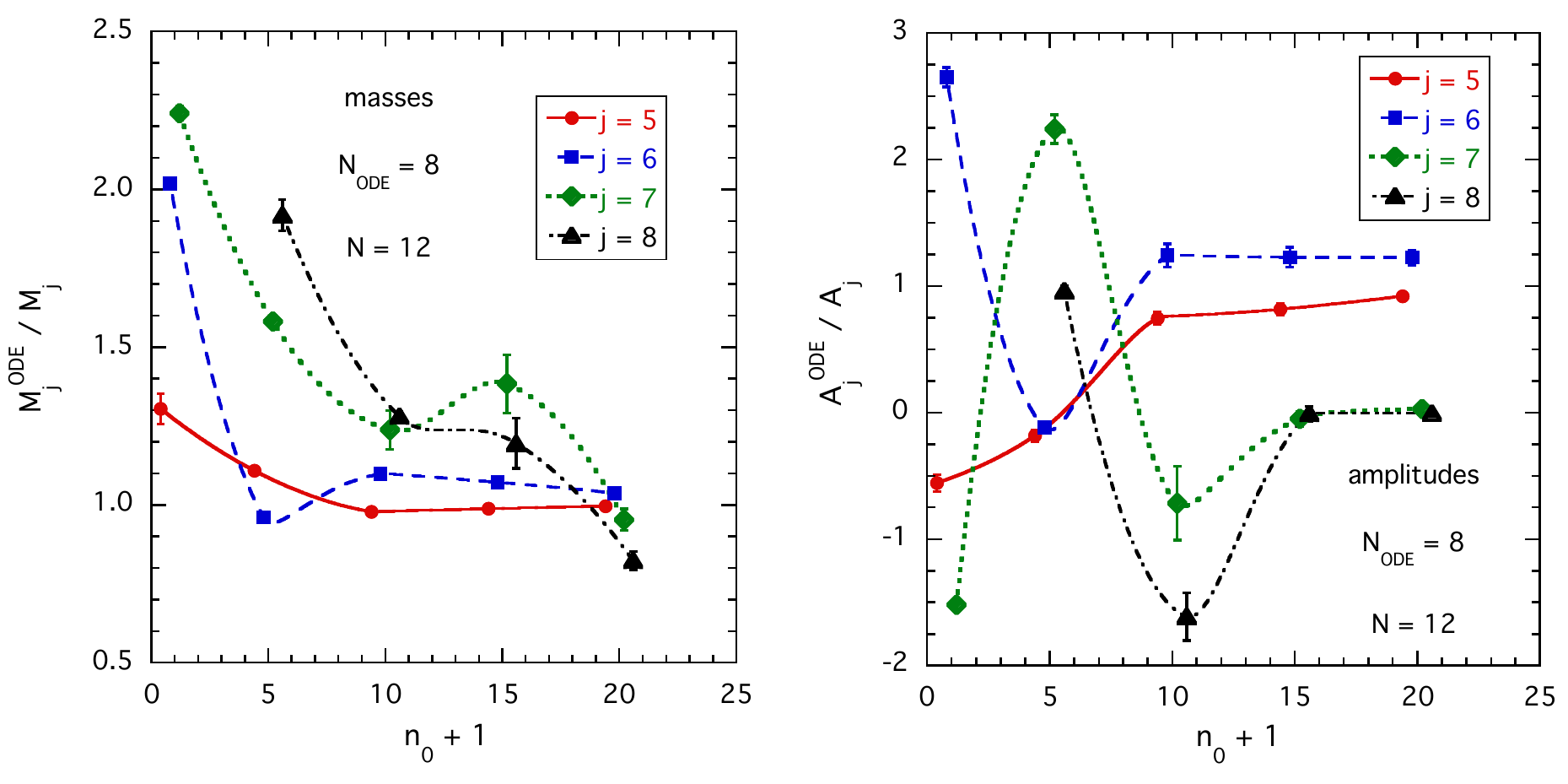}
\end{center}
\vspace{-0.75cm}
\caption{\it \small The same as in Fig.~\ref{fig:ODE8_1} but for the heavier four ODE states obtained assuming $N_{ODE} = 8$.}
\label{fig:ODE8_2}
\end{figure}

The remarkable stability of the masses and amplitudes of the lighter four states, determined by the ODE algorithm with $N_{ODE} = 8$, for various ranges of the analysis is also illustrated in Fig.~\ref{fig:ODE8_1}, where the results for the ratios $M_j^{ODE} / M_j$ and $A_j^{ODE} / A_j$ are shown.
On the contrary, an approximate stability is found for the fifth and sixth states only at large values of $n_0$, as shown in Fig.~\ref{fig:ODE8_2}. 
A very bad convergence of the masses and amplitudes of the heavier two ODE states ($j = 7$ and $j = 8$) is clearly visible.

\begin{figure}[htb!]
\begin{center}
\includegraphics[scale=0.85]{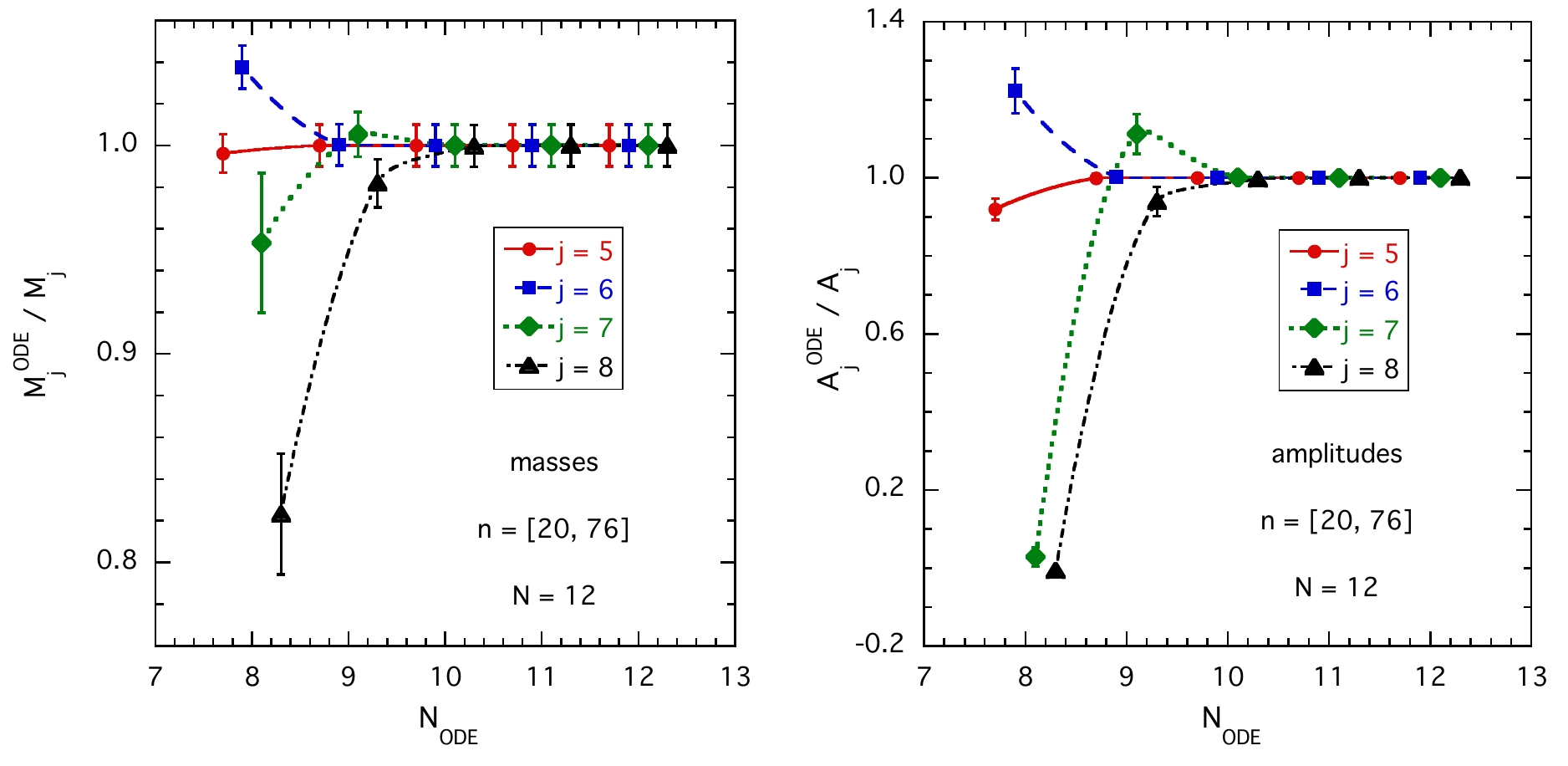}
\end{center}
\vspace{-0.75cm}
\caption{\it \small Ratios $M_j^{ODE} / M_j$ (left panel) and $A_j^{ODE} / A_j$ (right panel) of the ODE states with $j = 5, ..., 8$, obtained adopting the range of analysis $n = [20, \, 76]$, versus the total number $N_{ODE}$ of signals included in Eq.~(\ref{eq:Cn_ODE}), starting from $N_{ODE} = 8$ up to $N_{ODE} = N = 12$. The results for the various states are slightly shifted horizontally for better readability.}
\label{fig:ODE_20-76}
\end{figure}

The number of exponential signals properly reproduced by the ODE algorithm increases as $N_{ODE}$ increases.
As already pointed out in Section~\ref{sec:ODE}, this is due to the fact that in any fixed range of values of $n$ the derivative $C_n^{(2k)}$ is more sensitive to the signals with higher masses as the order $k$ increases (cf., e.g., the factor $(\overline{z}_i)^k$ in Eq.~(\ref{eq:C2kn})).
This feature is illustrated in Fig.~\ref{fig:ODE_20-76}, where for a given choice of the analysis range ($n = [20, \, 76]$) the convergence for the masses and amplitudes of the ODE states with $j = 5, .., 8$ toward the input values is clearly visible.

For sake of completeness we now generalize the structure of the ODE representation~(\ref{eq:Cn_ODE}) to the case of multiple exponential signals having masses $M_j^{ODE}$ with non-vanishing imaginary parts and multiplicity $\mu_j$ greater than $1$.
One has
\be
     C_n^{(ODE)} = \sum_{j=1}^{\overline{N}_{ODE}} \sum_{\mu=1}^{\mu_j} \overline{B}_{j \mu}^{ODE} ~ \overline{f}_n^{(j, \, \mu)} ~ ,
     \label{eq:Cn_ODE_mu}
\ee
where $\sum_{j=1}^{\overline{N}_{ODE}} \mu_j = N_{ODE}$ and
\be
    \overline{f}_n^{(j, \, \mu)} \equiv n^{\mu-1} s_n^j e^{- a {\rm Re}(M_j^{ODE}) n}  + (-)^p (N_T - n)^{\mu-1}  s_{N_T -n}^j e^{- a {\rm Re}(M_j^{ODE}) (N_T - n)} ~ .
    \label{eq:fn_j_Im_mu}
\ee
with $s_n^j$ given in Eq.~(\ref{eq:sn_j}) and $(-)^p$ being the $t$-parity of the correlator.

Finally, in the ODE procedure we introduce two tolerance factors $\delta_R$ and $\delta_I$, which help in finding multiple poles.
They are defined as follows.
When the real parts of the masses of two signals are close enough, namely
\be
    | a {\rm Re}(M_j^{ODE}) - a {\rm Re}(M_{j^\prime}^{ODE}) | < \delta_R \frac{2}{N_T} ~ ,
\ee
the two signals are replaced by a signal having as real part $[\mu_j a {\rm Re}(M_j^{ODE}) + \mu _{j^\prime} a {\rm Re}(M_{j^\prime}^{ODE})] / (\mu_j + \mu_{j^\prime})$ and multiplicity $\mu_j + \mu_{j^\prime}$.
When the imaginary part of the mass of a signal is small enough, namely
\be
    | a {\rm Im}(M_j^{ODE}) | < \delta_I \frac{2}{N_T} ~ ,
\ee
it is put to zero. 
By using the above tolerance factors the relative error in the ODE representation (\ref{eq:Cn_ODE_mu})) is of the order ${\cal{O}}(\delta^2)$.

The results presented in this Section illustrate that, even when the total number of exponential signals contained in the fake data is not known, the ODE method guarantees a quite good convergence toward accurate results for both masses and amplitudes, including their statistical fluctuations, at least for a significant subset of the exponential signals present in the fake correlator. 
We stress that the specific structure of the ODE representation [see Eq.~(\ref{eq:Cn_ODE}) or Eq.~(\ref{eq:Cn_ODE_mu})] does not require any {\it a priori} assumption, but it is properly detected by the ODE method.

\subsection{ODE analyses with $N_{ODE} > N$}
\label{sec:N_ODE+}

In Section~\ref{sec:fake_data} we have illustrated the nice performance of the ODE algorithm applied to the case in which the number of derivatives of the correlator included in the construction of the mass matrix is equal to the number of exponential signals included in the fake correlator, i.e.~$N_{ODE} = N$.
According to the numerical precision of the correlator the ODE algorithm detects almost exactly all the masses and the amplitudes (together with their statistical fluctuations) used as input (see Table~\ref{tab:Nmax}).

In this Section we describe an interesting feature of the ODE algorithm when $N_{ODE} > N$, namely when the number of roots of the polynomial $\overline{P}_{N_{ODE}}(z)$ (see Eq.~(\ref{eq:PNz_ODE})) is larger than the number $N$ of signals present in the correlator $C_n^{(0)}$.

Let us consider again the fake data for the correlator (\ref{eq:fake_Cn}) corresponding only to the first eight masses and amplitudes given in Table~\ref{tab:bench1} with $T / a = N_T = 96$.
When $N_{ODE} = N = 8$ the ODE algorithm determines precisely the input masses and amplitudes (together with their statistical fluctuations).
The absolute residue $\left| C_n^{(0)} - C_n^{(ODE)} \right|$ between the input correlator and its ODE representation is numerically very small ($< 10^{-20}$), and it corresponds to the level of the numerical rounding of the whole procedure.

We now apply the ODE algorithm using $N_{ODE} = 10, 12, 14, 16$.
The condition number of the mass matrix quickly increases up to $\approx 10^{70}$ for $N_{ODE} = 16$.
Therefore we adopt 96 digits for the internal ODE computations.
For all values $N_{ODE} > 8$ the eight roots of the fake correlator are properly determined, but also additional ($N_{ODE} - 8$) roots are found.
The interesting feature is that the values of $\overline{z}_j^{ODE} \equiv 2 \left[ \mbox{cosh}(a M_j^{ODE}) - 1 \right]$ for $j > 8$ are real and in the range $(-4, \, 0)$, which in turn mean imaginary values of the corresponding ODE masses $a M_j^{ODE}$, as shown in Table~\ref{tab:NODE+}.
\begin{table}[htb!]
\begin{center}
\begin{tabular}{||c|c|c||c|c||}
\hline
 $j$ & $N_{ODE} = 10$ & $N_{ODE} = 12$ & $N_{ODE} = 14$ & $N_{ODE} = 16$ \\
\hline \hline
 $~9$ & $i (2.873 \pm 0.495)$ & $i (2.960 \pm 0.477)$ & $i (2.996 \pm 0.446)$ & $i (3.023 \pm 0.479)$ \\
\hline
 $10$ & $i (2.312 \pm 0.561)$ & $i (2.576 \pm 0.428)$ & $i (2.722 \pm 0.472)$ & $i (2.804 \pm 0.433)$ \\
\hline
 $11$ &                                    & $i (2.211 \pm 0.431)$ & $i (2.430 \pm 0.472)$ & $i (2.535 \pm 0.439)$ \\
\hline
 $12$ &                                    & $i (1.799 \pm 0.486)$ & $i (2.138 \pm 0.432)$ & $i (2.317 \pm 0.487)$ \\
\hline 
$13$ &                                     &                                     & $i (1.822 \pm 0.375)$ & $i (2.089 \pm 0.437)$ \\
\hline
$14$ &                                     &                                     & $i (1.482 \pm 0.409)$ & $i (1.804 \pm 0.413)$ \\
\hline
$15$ &                                     &                                     &                                    & $i (1.554 \pm 0.430)$ \\
\hline
$16$ &                                     &                                      &                                   & $i (1.273 \pm 0.460)$ \\
\hline
\end{tabular}
\end{center}
\vspace{-0.25cm}
\caption{\it Masses $a M_j^{ODE}$ of the extra ($N_{ODE} - 8$) signals found by the ODE algorithm when $N_{ODE} = 10, 12, 14, 16$ is used for analyzing the correlator (\ref{eq:fake_Cn}), in which eight exponential signals, corresponding to the first eight masses and amplitudes given in Table~\ref{tab:bench1}, have been included.}
\label{tab:NODE+}
\end{table}
The extra ($N_{ODE} - 8$) roots correspond to {\bf noisy signals}, that the ODE algorithm includes for trying to take into account the small residual terms.
Note that the extra masses of Table~\ref{tab:NODE+} exhibit an approximate pattern, whose interpretation requires however a separate study.
The noisy roots can be simply discarded while keeping the remaining eight roots.
The ODE amplitudes $A_j^{ODE}$ corresponding to the non-noisy signals can be determined from the same amplitude matrix calculated when $N_{ODE} = N = 8$.

\section{Analysis of lattice correlators}
\label{sec:lattice}

In this Section we apply the ODE algorithm to selected correlation functions evaluated by means of large-scale QCD simulations on the lattice.
We will employ in particular correlators evaluated using gauge ensembles produced by the European (now Extended) Twisted Mass Collaboration (ETMC).
The numerical precision of the available correlators is optimistically the double precision (16 digits).
From Table~\ref{tab:Nmax} we expect that the ODE algorithm should detect accurately no more than 6 exponential signals.
However, in the case of lattice correlators the noise is not only due to the numerical rounding, but e.g.~also to the residues coming from gauge variant terms.
Therefore, we expect that only 3-4 exponential signals can be identified properly by the ODE algorithm.

Let's start by considering the 2-point correlator $C_{PS}(t)$ defined as
 \be
    C_{PS}(t) = \frac{1}{L^3} \sum\limits_{x, z} \left\langle 0 \right| P_5 (x) P_5^\dag (z) \left| 0 \right\rangle \delta_{t, (t_x  - t_z )} ~ ,
    \label{eq:P5P5}
 \ee
where $P_5 (x) = \overline{q}_2(x) \gamma_5 q_1(x)$ is a local interpolating field that creates in $x$ a pseudoscalar (PS) mesons made of two valence quarks $\bar{q}_1$ and $q_2$ with masses $m_1$ and $m_2$.
At large time distances one has
 \be
    C_{PS}(t)_{ ~ \overrightarrow{t  \gg a, ~ (T - t) \gg a} ~ } \frac{|\mathcal{Z}_{PS}|^2}{2M_{PS}} \left[ e^{ - M_{PS}  t}  + e^{ - M_{PS}  (T - t)} \right] ~ ,
    \label{eq:larget}
 \ee
so that the meson mass $M_{PS}$ and the matrix element $\mathcal{Z}_{PS} = \langle PS | \overline{q}_2 \gamma_5 q_1 | 0 \rangle$ can be extracted from the exponential fit given in the r.h.s.\,of Eq.~(\ref{eq:larget}). 

The evaluation of the correlator (\ref{eq:P5P5}) involves the determination of the so-called all-to-all quark propagator.
For the latter the statistical accuracy is significantly improved by using the so-called ``one-end" stochastic method~\cite{McNeile:2006bz}, which includes spatial stochastic sources at a single time slice chosen randomly.

Among the gauge ensembles generated by ETMC with $N_f = 2+1+1$ dynamical quarks (see Ref.~\cite{Carrasco:2014cwa}), we have selected the case of the gauge ensemble B25.32 corresponding to a lattice volume $V \times T = 32^3 \times 64 \, a^4$, to a lattice spacing $a \simeq 0.082$ fm with a light-quark mass equal to $m_\ell \simeq 12$ MeV (in the $\overline{\rm MS}(2~\mbox{GeV})$ scheme) and a strange and charm masses close to their physical values.
The number of independent gauge configurations employed is 150.
The correlator (\ref{eq:P5P5}) has been calculated using 160 stochastic sources (diagonal in the spin variable and dense in the color one) per gauge configuration for valence quark masses equal to $m_1 = m_2 = m_\ell$, corresponding to a simulated pion mass of $\simeq 260$ MeV (see Ref.~\cite{Giusti:2018mdh}).
The relative statistical error $\sigma_{PS}(t) / C_{PS}(t)$ does not exceed $\simeq 1.4 \%$. 

\begin{figure}[htb!]
\begin{center}
\includegraphics[scale=0.415]{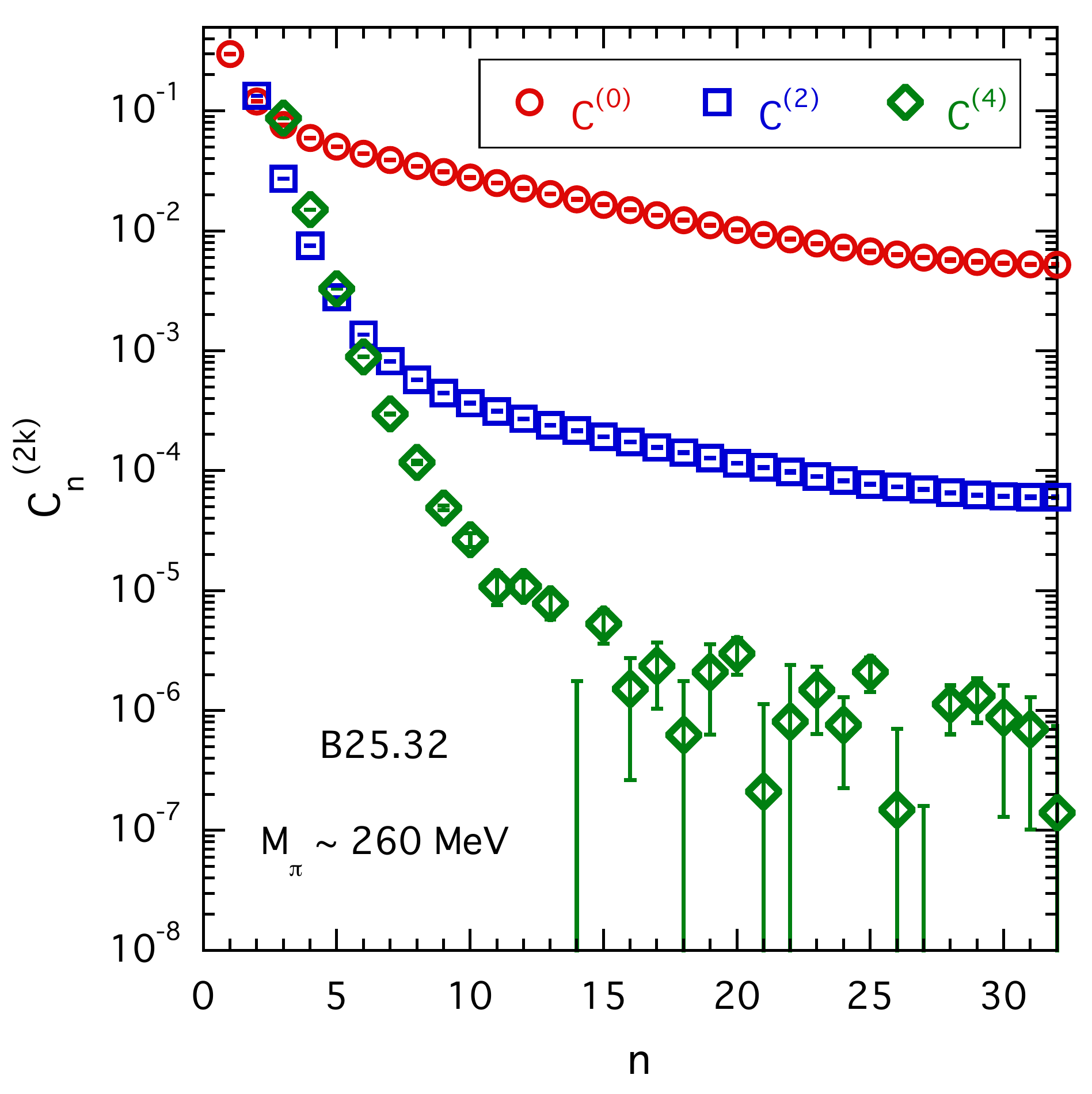}
\includegraphics[scale=0.415]{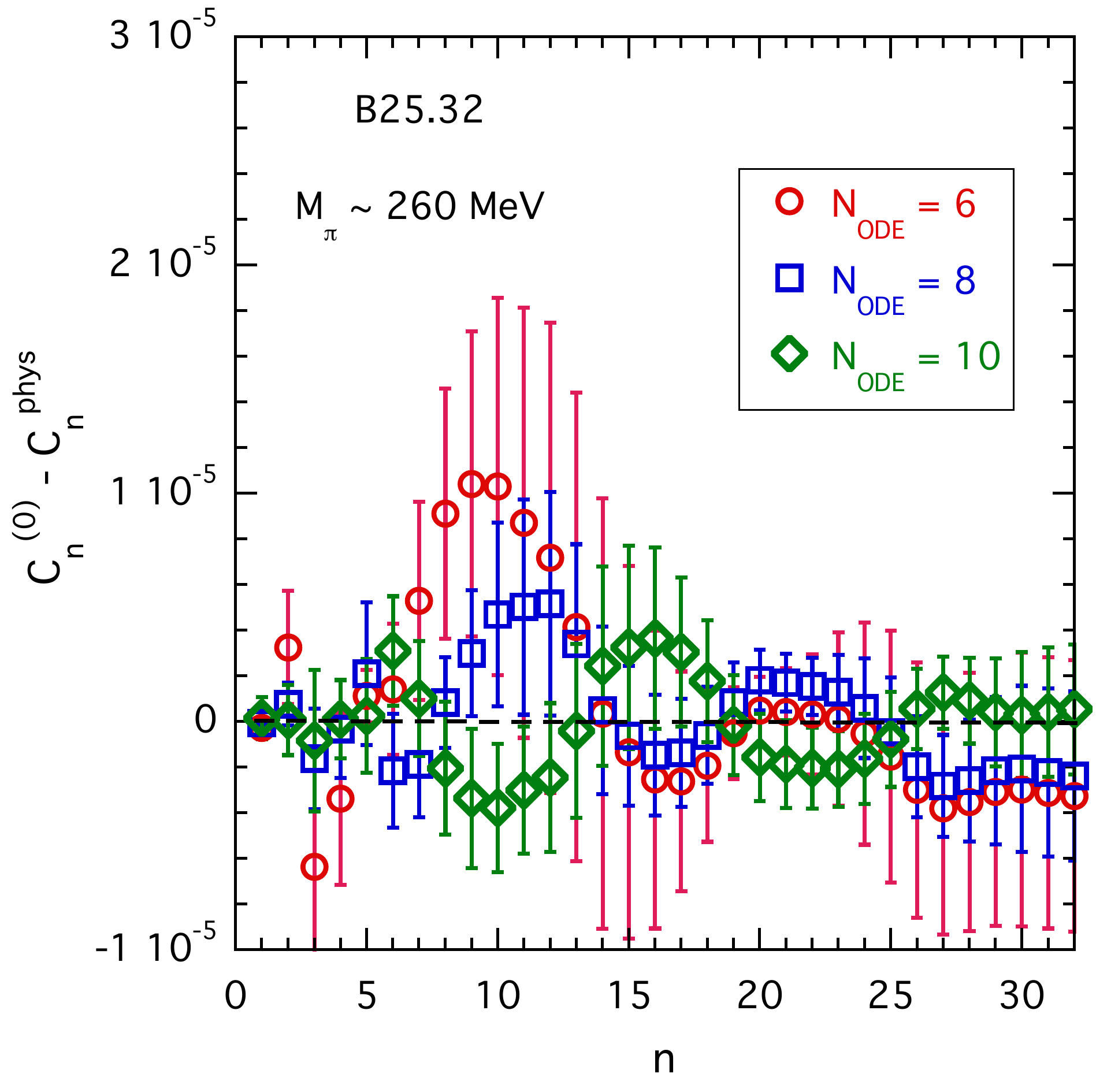}
\end{center}
\vspace{-1cm}
\caption{\it \small Left panel: time dependencies of the correlator (\ref{eq:P5P5}) and its second and fourth derivatives, $C_n^{(2k)}$ with $k = 0, 1, 2$, corresponding to the ETMC gauge ensemble B25.32 (see text). Right panel: time dependence of the difference $C_n^{(0)} - C_n^{phys}$, where $C_n^{phys}$ is the ODE representation (\ref{eq:Cn_phys}) (see later on) corresponding to $N_{ODE} = 6, 8, 10$.}
\label{fig:B2532}
\end{figure}

The ETMC correlator (\ref{eq:P5P5}) and its second and fourth derivatives are shown in the left panel of Fig.~\ref{fig:B2532}.
With respect to the case of fake data (see the left panel of Fig.~\ref{fig:bench1}) the fourth derivative (as well as higher ones) is manifestly noisy at large time distances.

We have applied the ODE algorithm using values of $N_{ODE}$ from $4$ up to $10$ in the full range of values of $n$, i.e.\,$n = [1, \, 64]$.
In the case $N_{ODE} = 4$ we find three exponential signals, which are non-noisy and non-oscillating signals (i.e.~$\overline{z}_j^{ODE} > 0$), and $1$ noisy signal of the type described in the previous Section (i.e.~$-4 < \overline{z}_j^{ODE} < 0$).

In what follows we refer to the non-noisy states as the {\bf physical states}.

When $N_{ODE} > 4$ we find always four physical signals and $N_{ODE} - 4$ noisy states.
For all values of $N_{ODE}$ physical and noisy states have multiplicity equal to $1$.
Thus, for the correlator (\ref{eq:P5P5}) the ODE representation corresponding only to the physical states is given by
\be
    C_n^{phys} = \sum_{j=1}^{N_{phys}} A_j^{ODE} ~ \left[ e^{-aM_j^{ODE} n } + e^{-aM_j^{ODE} (N_T - n) } \right] ~ ,
    \label{eq:Cn_phys}
\ee
where $N_{phys}$ indicates the number of physical states.
The difference $C_{PS}(t) - C^{phys}(t)$ is therefore related to the noisy states only and it is shown in the right panel of Fig.~\ref{fig:B2532} for $N_{ODE} = 6, 8, 10$.
The noisy states interfere clearly with the detection of excited states and the number of physical exponential signals, that can be extracted by the ODE method, is therefore limited by the level of the noise.
We speculate that the observed noise can be reduced by increasing the number of gauge configurations used.
This point requires, however, further dedicated investigations.

The results for the physical signals obtained with $N_{ODE} = 4$, $5$, $6$, $8$ and $10$ are collected in Table~\ref{tab:B2532_160}.
\begin{table}[htb!]
{\small
\begin{center}
\begin{tabular}{||c||c|c|c|c|c||}
\hline
 $j$ & $a M_j^{ODE}$    & $a M_j^{ODE}$   & $a M_j^{ODE}$    & $a M_j^{ODE}$   & $a M_j^{ODE}$\\
       & $(N_{ODE} = 4)$ & $(N_{ODE} = 5)$ & $(N_{ODE} = 6)$ & $(N_{ODE} = 8)$ & $(N_{ODE} = 10)$\\
 \hline \hline
 $1$ & $0.10935 \pm 0.00088$ & $0.10644 \pm 0.00075$ & $0.10693 \pm 0.00027$ & $0.10694 \pm 0.00034$ & $0.10697 \pm 0.00034$  \\
\hline
 $2$ & $\,0.9625 \pm 0.0452\,$ & $\,0.5126 \pm 0.0903\,$ & $\,0.5839 \pm 0.0307\,$ & $\,0.5600 \pm 0.0463\,$ & $\,0.5872 \pm 0.0432\,$ \\
\hline
 $3$ & $\,2.0201 \pm 0.0311\,$ & $\,1.1453 \pm 0.0931\,$ & $\,1.2147 \pm 0.0586\,$ & $\,1.1570 \pm 0.0920\,$ & $\,1.2168 \pm 0.1004\,$   \\
\hline
 $4$ &                                        & $\,2.0910 \pm 0.0568\,$ & $\,2.1321 \pm 0.0427\,$ & $\,2.0888 \pm 0.0651\,$ & $\,2.1311 \pm 0.0819\,$   \\
\hline
\end{tabular}
\vskip 0.5cm
\begin{tabular}{||c||c|c|c|c|c||}
\hline
$j$ & $A_j^{ODE}$       & $A_j^{ODE}$       & $A_j^{ODE}$       & $A_j^{ODE}$        & $A_j^{ODE}$\\
      & $(N_{ODE} = 4)$ & $(N_{ODE} = 5)$ & $(N_{ODE} = 6)$ & $(N_{ODE} = 8)$ & $(N_{ODE} = 10)$\\
 \hline \hline
 $1$ & $0.08369 \pm 0.00093$ & $0.07934 \pm 0.00085$ & $0.08015 \pm 0.00058$ & $0.08014 \pm 0.00055$ & $0.08021 \pm 0.00054$ \\
\hline
 $2$ & $\,0.2442 \pm 0.0271\,$ & $\,0.0347 \pm 0.0140\,$ & $\,0.0463 \pm 0.0090\,$ & $\,0.0387 \pm 0.0125\,$ & $\,0.0468 \pm 0.0137\,$ \\
\hline
 $3$ & $\,0.9809 \pm 0.0221\,$ & $\,0.2936 \pm 0.0454\,$ & $\,0.3272 \pm 0.0342\,$ & $\,0.2938 \pm 0.0504\,$ & $\,0.3281 \pm 0.0659\,$  \\
\hline
 $4$ &                                        & $\,0.9142 \pm 0.0520\,$ & $\,0.8723 \pm 0.0350\,$ & $\,0.9040 \pm 0.0489\,$ & $\,0.8693 \pm 0.0641\,$  \\
 \hline
\end{tabular}
\end{center}
}
\vspace{-0.25cm}
\caption{\it Masses $a M_j^{ODE}$ (upper panel) and amplitudes $A_j^{ODE}$ (lower panel) of the four physical signals found by the ODE algorithm when $N_{ODE} = 4$, $5$, $6$, $8$ and $10$ is used for analyzing the PS correlator (\ref{eq:P5P5}) calculated in the case of the ETMC gauge ensemble B25.32~\cite{Carrasco:2014cwa} for valence quark masses equal to $m_1 = m_2 = m_\ell \simeq 12$ MeV in the $\overline{\rm MS}(2~\mbox{GeV})$ scheme.}
\label{tab:B2532_160}
\end{table}
A good convergence for the ground state and progressively for the excited states is observed as $N_{ODE}$ increases.
For $N_{ODE} \geq 6$ a nice agreement for both masses and amplitudes occurs within the uncertainties.
The latter ones appear to be slightly larger as $N_{ODE}$ increases above $N_{ODE} = 6$ and this is probably due to the occurrence of more extra roots in the mass matrix.
Note the strong stability of the mass and amplitude for the ground-state signal for $N_{ODE} \geq 6$. 

\begin{figure}[htb!]
\begin{center}
\includegraphics[scale=0.75]{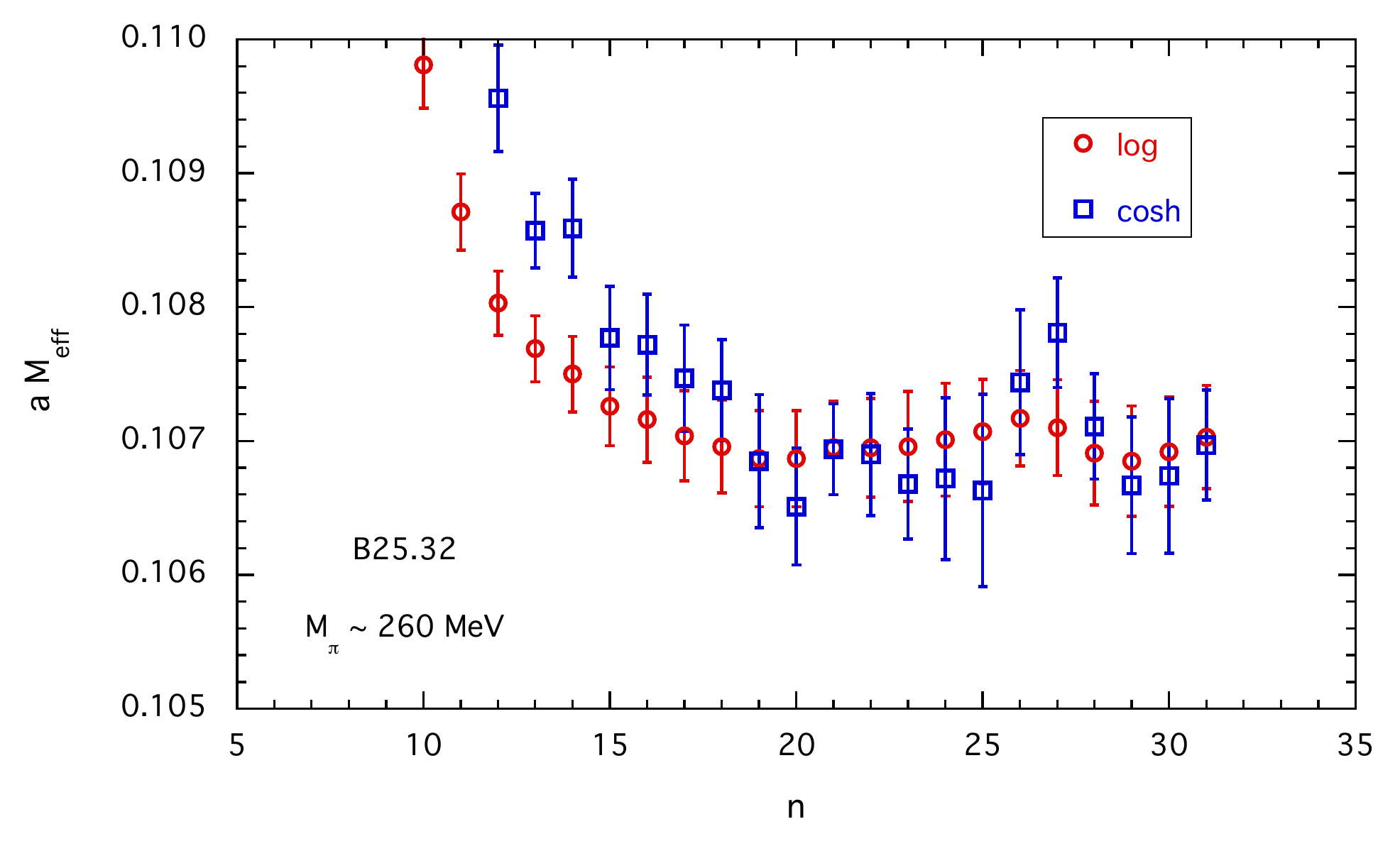}
\end{center}
\vspace{-0.75cm}
\caption{\it \small Time dependence of two definitions of the effective mass given respectively by Eqs.~(\ref{eq:Meff_cosh}) and (\ref{eq:Meff_log}) in the case of the PS correlator (\ref{eq:P5P5}) corresponding to a pion with mass $M_\pi \simeq 260$ MeV in the case of the ETMC gauge ensemble B25.32~\cite{Carrasco:2014cwa}.}
\label{fig:Meff_B2532}
\end{figure}

The approach commonly used to extract the ground-state signal is to look at the time dependence of the {\it effective} mass in order to identify the range of values of $n$, where the ground-state dominates.
In Fig.~\ref{fig:Meff_B2532} the results corresponding to two definitions of the effective mass are shown. 
One definition is given by Eq.~(\ref{eq:Meff_cosh}) and the other one is the usual logarithmic slope
\be
    a M_{eff}^{log} \equiv \mbox{log}\left( \frac{C_{n-1}^{(0)}}{C_n^{(0)}} \right) ~ .
    \label{eq:Meff_log}
\ee
At enough large time distances both definitions provide the mass of the ground-state. 
However, at lower values of $n$ they differ due the different impact of the excited states.
A single exponential fit in a chosen range $n = [n_{min}, \, n_{max}]$ provides values for $aM_1$ and $A_1$, which in general may depend on the choice of $n_{min}$ and $n_{max}$. 
Such a dependence may represent a possible source of a systematic uncertainty.
In the case at hand a single exponential fit in the range $[18, \, 30]$ yields $a M_1 = 0.10699~(37)$ and $A_1 = 0.08025~(55)$ in nice agreement with the corresponding ODE results of Table~\ref{tab:B2532_160}. 
The dependence of the chosen range turns out to be a sub-leading effect with respect to the statistical error.

We stress that the ODE procedure allows to extract the ground-state signal taking into account the presence of excited states without the need of finding a range of values of $n$, where the ground-state dominates.
The systematic uncertainty related to the choice of such a range (i.e.~to the contamination of excited states) is therefore avoided adopting the ODE procedure.
Moreover, important pieces of information on the excited states can be obtained for spectroscopic studies.

We now move to the heavier sector of charmonium.
We select the gauge ensemble D20.48~\cite{Carrasco:2014cwa} corresponding to a lattice volume $V \times T = 48^3 \times 96 a^4$ and to a lattice spacing $a \simeq 0.062$ fm, and choose the valence quark masses equal to $m_1 = m_2 \simeq 1.18$ GeV (in the $\overline{\rm MS}(2~\mbox{GeV})$ scheme), i.e.~close to the physical charm quark mass~\cite{Carrasco:2014cwa}.
The number of independent gauge configurations employed is 100.
We consider both the PS correlator (\ref{eq:P5P5}) and the vector (V) one given by
 \be
    C_V(t) = \frac{1}{L^3} \sum\limits_{x, z} \frac{1}{3} \sum_{i=1}^3 \left\langle 0 \right| V_i (x) V_i^\dag (z) \left| 0 \right\rangle \delta_{t, (t_x  - t_z )} ~ ,
    \label{eq:ViVi}
 \ee
where $V_i (x) = \overline{q}_2(x) \gamma_i q_1(x)$ is a local interpolating field that creates a vector meson in $x$.
Four stochastic sources per gauge configuration are employed leading to a relative statistical error which does not exceed $\sim 0.4 \%$ and $\sim 0.7 \%$ for the PS and V correlators, respectively.

We apply the ODE algorithm with $N_{ODE} \geq 4$ finding always four physical signals and $N_{ODE} - 4$ noisy states (all with multiplicity equal to $1$).
The results for the physical signals obtained with $N_{ODE} = 4$, $6$ and $8$ are collected in Tables~\ref{tab:D2048_cc_PS} and~\ref{tab:D2048_cc_VV} for the PS and V correlators, respectively.
The convergence of both masses and amplitudes for all the four physical states is quite good.
\begin{table}[htb!]
\begin{center}
\begin{tabular}{||c|||c|c|c||}
\hline
 $j$ & $a M_j^{ODE}$    & $a M_j^{ODE}$    & $a M_j^{ODE}$\\
       & $(N_{ODE} = 4)$ & $(N_{ODE} = 6)$ & $(N_{ODE} = 8)$\\
 \hline \hline
 $1$ & $0.95201 \pm 0.00015$ & $0.95192 \pm 0.00015$ & $0.95189 \pm 0.00019$ \\
\hline
 $2$ & $\,1.2466 \pm 0.0054\,$ & $\,1.2348 \pm 0.0035\,$ & $\,1.2304 \pm 0.0051\,$ \\
\hline
 $3$ & $\,1.7938 \pm 0.0121\,$ & $\,1.7589 \pm 0.0093\,$ & $\,1.7480 \pm 0.0130\,$ \\
\hline
 $4$ & $\,2.6731 \pm 0.0284\,$ & $\,2.5969 \pm 0.0213\,$ & $\,2.5784 \pm 0.0247\,$ \\
\hline
\end{tabular}
\vskip 0.5cm
\begin{tabular}{||c|||c|c|c||}
\hline
$j$ & $A_j^{ODE}$       & $A_j^{ODE}$       & $A_j^{ODE}$\\
      & $(N_{ODE} = 4)$ & $(N_{ODE} = 6)$ & $(N_{ODE} = 8)$\\
 \hline \hline
 $1$ & $0.06477 \pm 0.00026$ & $0.06456 \pm 0.00023$ & $0.06448 \pm 0.00029$\\
\hline
 $2$ & $\,0.1198\pm 0.0042\,$ & $\,0.1093 \pm 0.0026\,$ & $\,0.1057 \pm 0.0037\,$ \\
\hline
 $3$ & $\,0.5438 \pm 0.0109\,$ & $\,0.5119 \pm 0.0100\,$ & $\,0.5036 \pm 0.0124\,$ \\
\hline
 $4$ & $\,0.4369 \pm 0.0118\,$ & $\,0.4717 \pm 0.0100\,$ & $\,0.4819 \pm 0.0138\,$ \\
 \hline
\end{tabular}
\end{center}
\vspace{-0.25cm}
\caption{\it Masses $a M_j^{ODE}$ (upper panel) and amplitudes $A_j^{ODE}$ (lower panel) of the four physical signals found by the ODE algorithm when $N_{ODE} = 4$, $6$ and $8$  is used for analyzing the PS correlator (\ref{eq:P5P5}) calculated in the case of the ETMC gauge ensemble D20.48~\cite{Carrasco:2014cwa} for valence quark masses in the charm region, namely $m_1 = m_2 \simeq 1.18$ GeV in the $\overline{\rm MS}(2~\mbox{GeV})$ scheme.}
\label{tab:D2048_cc_PS}
\end{table}
\begin{table}[htb!]
\begin{center}
\begin{tabular}{||c|||c|c|c||}
\hline
 $j$ & $a M_j^{ODE}$    & $a M_j^{ODE}$    & $a M_j^{ODE}$\\
       & $(N_{ODE} = 4)$ & $(N_{ODE} = 6)$ & $(N_{ODE} = 8)$\\
 \hline \hline
 $1$ & $0.99431 \pm 0.00104$ & $0.99556 \pm 0.00027$ & $0.99550 \pm 0.00022$ \\
\hline
 $2$ & $\,1.2458 \pm 0.0243\,$ & $\,1.2706 \pm 0.0070\,$ & $\,1.2689 \pm 0.0058\,$ \\
\hline
 $3$ & $\,1.8276 \pm 0.0330\,$ & $\,1.8472 \pm 0.0153\,$ & $\,1.8431 \pm 0.0134\,$ \\
\hline
 $4$ & $\,2.7898 \pm 0.0354\,$ & $\,2.7943 \pm 0.0217\,$ & $\,2.7892 \pm 0.0197\,$ \\
\hline
\end{tabular}
\vskip 0.5cm
\begin{tabular}{||c|||c|c|c||}
\hline
$j$ & $A_j^{ODE}$       & $A_j^{ODE}$       & $A_j^{ODE}$\\
      & $(N_{ODE} = 4)$ & $(N_{ODE} = 6)$ & $(N_{ODE} = 8)$\\
 \hline \hline
 $1$ & $0.01615 \pm 0.00051$ & $0.01675 \pm 0.00014$ & $0.01672 \pm 0.00012$\\
\hline
 $2$ & $\,0.0341\pm 0.0040\,$ & $\,0.0378 \pm 0.0015\,$ & $\,0.0374 \pm 0.0013\,$ \\
\hline
 $3$ & $\,0.2104 \pm 0.0098\,$ & $\,0.2109 \pm 0.0061\,$ & $\,0.2094 \pm 0.0054\,$ \\
\hline
 $4$ & $\,0.3630 \pm 0.0113\,$ & $\,0.3577 \pm 0.0058\,$ & $\,0.3592 \pm 0.0048\,$ \\
 \hline
\end{tabular}
\end{center}
\vspace{-0.25cm}
\caption{\it The same as in Table~\ref{tab:D2048_cc_PS} but in the case of the vector correlator (\ref{eq:ViVi}).}
\label{tab:D2048_cc_VV}
\end{table}

\begin{figure}[htb!]
\begin{center}
\includegraphics[scale=0.415]{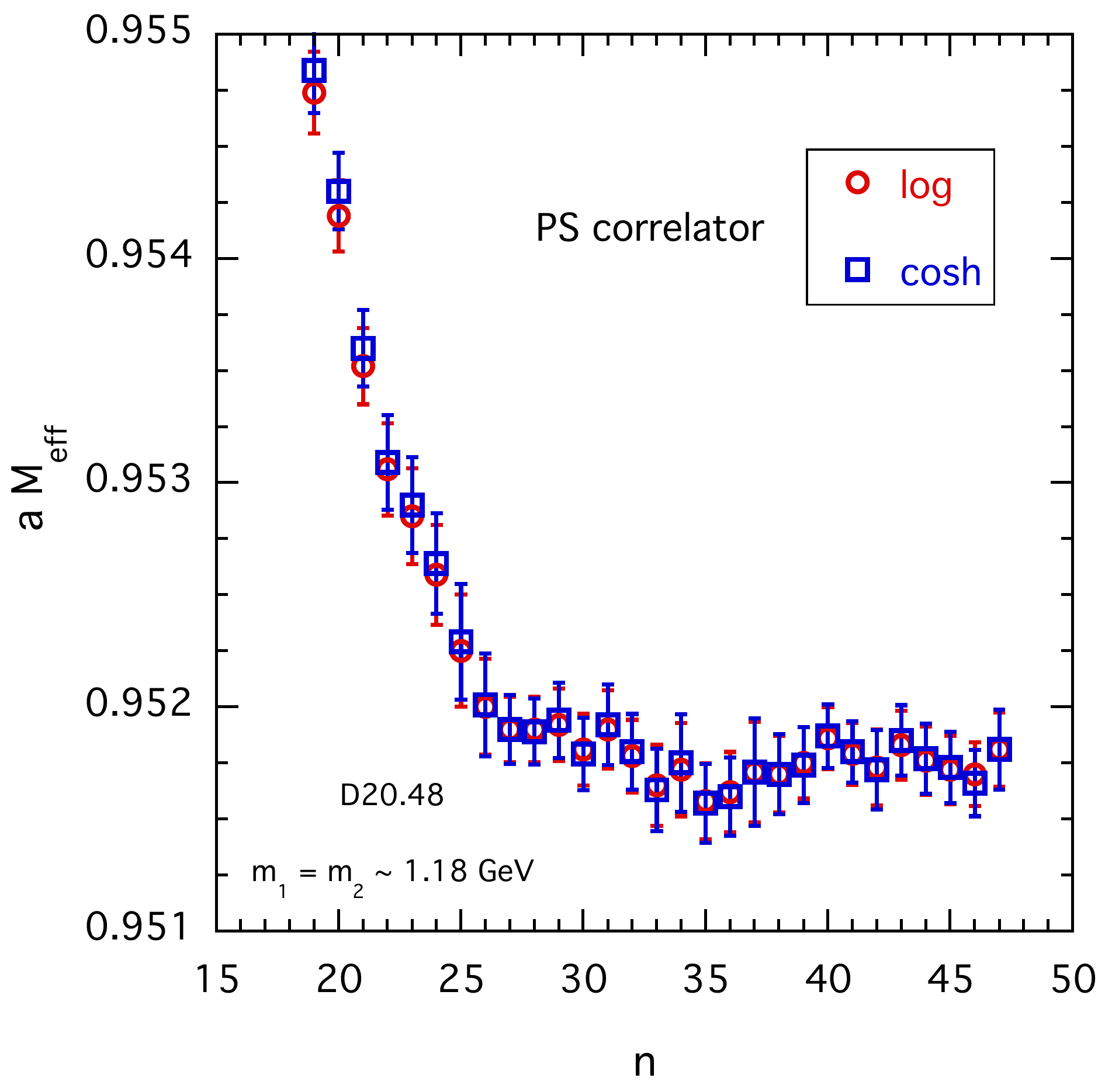}
\includegraphics[scale=0.415]{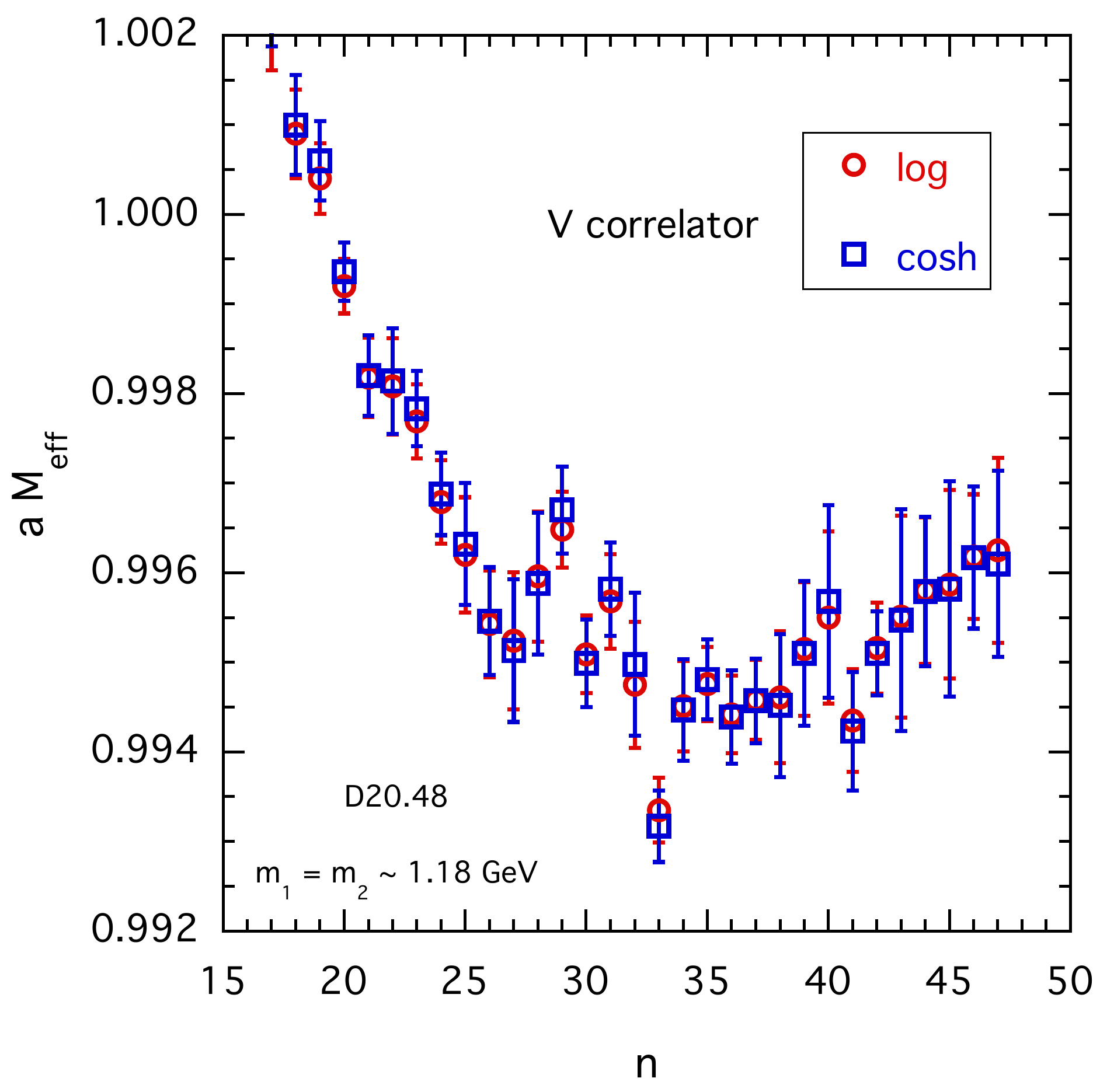}
\end{center}
\vspace{-0.75cm}
\caption{\it \small Time dependence of the effective mass (\ref{eq:Meff_cosh}) and (\ref{eq:Meff_log}) in the case of the PS (left panel) and V (right panel) correlators corresponding to two mass-degenerate valence quarks in the charm region in the case of the ETMC gauge ensemble D20.48~\cite{Carrasco:2014cwa}.}
\label{fig:Meff_D2048}
\end{figure}

The time dependence of the effective masses (\ref{eq:Meff_cosh}) and (\ref{eq:Meff_log}) is shown in Fig.~\ref{fig:Meff_D2048}.
In the PS case a single exponential fit in the range $n = [26,\,44]$ provides the values $aM_1 = 0.95175~(14)$ and $A_1= 0.06426~(25)$, which agree well with the corresponding ODE results shown in Table~\ref{tab:D2048_cc_PS}.

In the V case the plateau of the effective mass is of lower quality and the dependence of the single exponential fit on the chosen range $[n_{min}, \, n_{max}]$ is more relevant, as it is shown in Table~\ref{tab:Meff_D2048_VV}.
\begin{table}[htb!]
\begin{center}
\begin{tabular}{||c||c|c||}
\hline
$[n_{min}, \, n_{max}]$ & $aM_1$ & $A_1$ \\
 \hline \hline
 $[25, \, 45]$ & $0.99492 \pm 0.00018$ & $0.01642 \pm 0.00013$ \\
\hline
 $[30, \, 45]$ & $0.99463 \pm 0.00024$ & $0.01625 \pm 0.00018$ \\
\hline
 $[25, \, 40]$ & $0.99494 \pm 0.00020$ & $0.01643 \pm 0.00013$ \\
\hline
 $[30, \, 40]$ & $0.99460 \pm 0.00022$ & $0.01623 \pm 0.00018$ \\
\hline
\end{tabular}
\end{center}
\vspace{-0.25cm}
\caption{\it Ground-state mass, $aM_1$, and amplitude, $A_1$, extracted from a single exponential fit in a chosen range $n = [n_{min}, \, n_{max}]$ in the case of the V correlator (\ref{eq:ViVi}) corresponding to two valence quarks with masses $m_1 = m_2 \simeq 1.18$ GeV in the case of the ETMC gauge ensemble D20.48~\cite{Carrasco:2014cwa}.}
\label{tab:Meff_D2048_VV}
\end{table}
The corresponding systematic uncertainty turns out to be not negligible with respect to the statistical error.
Note also that the results of the single exponential fit lie below the corresponding ODE results, shown in Table~\ref{tab:D2048_cc_VV}, by approximately $2$ standard deviations.

\section{Use of the ODE algorithm together with other techniques}
\label{sec:improvement}

The results presented in the previous Sections show that the use of the ODE algorithm allows to extract multiple exponential signals from the temporal dependence of correlation functions without the need of any further input other than the correlator itself.
In particular, the ODE method is able to detect the multiplicities of the signals as well as the presence of possible oscillating signals.
This represents a relevant piece of information about the specific structure of the temporal dependence of the correlator. 

In this way even a quite involved representation, like the one given by Eq.~(\ref{eq:Cn_ODE_mu}), can be explicitly written down and the ODE algorithm provides values for both masses and amplitudes as well as their uncertainties.
All that can be used as a starting point for the the application of other techniques suitable for refining the fitting procedure of the temporal dependence of the correlator.

The simplest choice is to adopt a nonlinear least-squares minimizer starting from the ODE solution corresponding to the physical states.
In other words, the ODE physical states provide the specific structure of the temporal dependence of the correlator, where masses and amplitudes can be used as free parameters to be varied starting from the values obtained by the ODE algorithm and to be determined by minimizing a $\chi^2$-variable.
If the quality of the ODE representation of the correlator is adequate, then the subsequent application of the least-squares minimizer will either confirm the ODE values within the uncertainties or possibly refine the ODE solution.

Note that the ODE method does not correspond to the minimization of a unique $\chi^2$-variable. 
Indeed, as pointed out in Section~\ref{sec:ODE}, the inversion of the mass matrix is equivalent to minimize the variable $\chi_M^2$ defined by Eq.~(\ref{eq:chi2_mass}), while, once the masses are given, the determination of the amplitudes corresponds to the minimization of the $\chi^2$-variable given by Eqs.~(\ref{eq:chi2_N}-\ref{eq:chi2_k}) (or just its first term with $k=0$).
Furthermore, as shown in the previous Section, the ODE algorithm is sensitive to the presence of noisy states only through the mass matrix and independently of the size of the corresponding amplitudes.
Instead, a $\chi^2$-minimization procedure is expected to be sensitive also to the amplitudes of the noise.
In this way the combined ODE plus $\chi^2$-minimization procedure can produce either a non-trivial check of the ODE solution or possibly a refinement.  

We have explicitly used the above combination in the case of the lattice correlators analyzed in Section~\ref{sec:lattice}.
We have found that the ODE solution is nicely confirmed, within the uncertainties, by the subsequent $\chi^2$-minimization procedure.
An interesting feature is that no priors on the masses and amplitudes are required in the $\chi^2$ minimization in order to obtain stable results.
In the constrained curve fitting method of Ref.~\cite{Lepage:2001ym} priors are instead introduced just for stabilizing the results of the fitting procedure.

The combination of the ODE method with nonlinear least-squares minimizers requires, however, dedicated numerical investigations, which are outside the scope of the present paper.

We close this Section by observing that the number of hadronic states in a lattice QCD correlator (i.e.~the number of exponential signals) is not finite and, generally speaking, it increases as the time distance decreases.
In this respect it is worth to mention the representation of the vector-vector current correlator, relevant for the determination of the hadronic contribution to the anomalous magnetic moment of the muon, developed using quark-hadron duality at short time distances in Ref.~\cite{Giusti:2018mdh}.
The {\it dual} contribution represents an effective way to perform a resummation of an infinite number of highly excited hadronic states at short time distances.
Thus, the ODE algorithm can be combined with quark-hadron duality: it can be applied not to the full correlator, but to its difference with the {\it dual} contribution.
Such an issue will be investigated in a separate work.

\section{Conclusions}
\label{sec:conclusions}

We have presented a fast and simple algorithm that allows the extraction of multiple exponential signals from the temporal dependence of correlation functions evaluated on the lattice including the statistical fluctuations of each signal and treating properly backward signals.

The method starts from well-known features of the solution of ordinary (linear) differential equations and extracts multiple exponential signals from a generic correlation function simply by inverting appropriate mass and amplitude matrices and by finding the roots of an appropriate polynomial.
The method is based on the use of discretized derivatives of the correlation function.
An important feature of the ODE method is the level of singularity of the mass matrix, described by its condition number. 
On one hand side this represents a positive feature, that allows the ODE algorithm to be sensitive to the fluctuations of the exponential signals, and on the other hand side it is a limiting factor related to the presence of noise in the correlation function.
Huge values of the condition number require an appropriate treatment of the numerical precision, which has been obtained adopting the multiple precision software from Ref.~\cite{mpfun}.
Moreover, the root finding has been carried out accurately using the open-source software package MPSolve~\cite{MPSolve}.

We have tested extensively the ODE method using fake data, generated assuming a fixed number of exponential signals included in the correlator with a controlled numerical precision and within given statistical fluctuations.
All the exponential signals with their statistical uncertainties are determined exactly by the ODE algorithm, when the total number of exponential signals is known.
The only limiting factor is the numerical rounding off.
We have shown that, even when the total number of exponential signals contained in the correlator is not known, the ODE method guarantees a quite good convergence toward accurate results for both masses and amplitudes, including their statistical fluctuations, at least for a significant subset of the exponential signals present in the correlator.

Then, few cases of correlation functions evaluated by means of large-scale QCD simulations on the lattice have been addressed explicitly. 
In the case of lattice correlators, several sources of noise, other than the numerical rounding, can affect the correlator.
As shown in Section~\ref{sec:lattice}, the noise represents the crucial factor limiting the number of physical exponential signals, related to the hadronic spectral decomposition of the correlation function, that can be determined.

We have illustrated that the ODE algorithm can be applied to a large variety of correlation functions typically encountered in QCD or QCD+QED simulations on the lattice, including the case of exponential signals corresponding to poles with arbitrary multiplicity and/or the case of oscillating signals. 
Two important features of the ODE algorithm are: ~ i) its ability to detect the proper structure of the multiple exponential signals without any {\it a priori} assumption, and ~ ii) the extraction of the ground-state signal with accuracy without the need that the lattice temporal extension is large enough to allow the ground-state signal to be isolated.
This is a very useful property, which in particular can take care properly of the contamination of the excited states in the lattice correlators used for the determination of hadronic quantities, like e.g.~the form factors.

A further application of the ODE algorithm is represented by its combination with a subsequent nonlinear least-squares minimizer, where masses and amplitudes are used as free parameters (without any prior) to be varied starting from the values obtained by the ODE method.

A careful study of the origin and the structure of the noisy states, which are systematically detected by the ODE algorithm when analyzing lattice correlators, requires dedicated numerical investigations, that we are planning.
At the same time applications of the ODE algorithm to the analysis of 2-point and 3-point correlation functions evaluated on the lattice are in progress.

\section*{Acknowledgments}

We gratefully acknowledge V.~Lubicz and F.~Sanfilippo for fruitful discussions and D.~Giusti for a careful reading of the manuscript.
S.S.~warmly thanks D.H.~Bailey for his valuable assistance in the use of the MPFUN2015 software package.

\appendix

\section{Filtering of signals}
\label{sec:appendix}

In this Appendix we make use of the ODE algorithm to address the issue of subtracting exponential signals from a given correlator without the need of determining their amplitudes.

Without loss of generality let's consider the case of a correlator with a given $t$-parity $(-)^p$, composed by $N$ exponential signals each with multiplicity equal to $1$, namely
\be
     C_n^{(0)} = \sum_{i=1}^N A_i \left[ e^{- a M_i n} + (-)^p e^{- a M_i (N_T - n)} \right] ~ .
     \label{eq:Cn_initial}
\ee
The first step of the ODE algorithm is to calculate iteratively the derivatives 
\be
   C_n^{(2k)} = C_{n+1}^{(2k-2)} + C_{n-1}^{(2k-2)} - 2 C_n^{(2k-2)} 
   \label{eq:C2kn_initial}
\ee
for $k = 1, ..., N$ and construct the mass matrix (\ref{eq:mass_matrix_p}) and the vector (\ref{eq:vectorV_p}).
Then, the linear system of equations (\ref{eq:mass_eq_p}) can be solved numerically to obtain the coefficients $x_k$, which define the polynomial
\be
    P_N(z) =  \sum_{k=0}^{N-1} x_k z^k + z^N = \prod_{i=1}^N (z - z_i) ~ 
     \label{eq:PNz_initial}
\ee
having $N$ roots located at $z_i = 2 \left[ \mbox{cosh}(a M_i) - 1 \right]$.

Now we want to subtract a subset of exponential signals, for instance $M$ exponentials having the masses $M_i$ with $i = 1, ..., M$.
First of all, we rewrite the polynomial (\ref{eq:PNz_initial}) as
\be
      P_N(z) =  Q_M(z) \cdot R_{N-M}(z)
      \label{eq:product}
\ee
where
\bea
      \label{eq:QMz}
       Q_M(z) & \equiv &  \prod_{i=1}^M (z - z_i) ~ , ~ \\[2mm]
      \label{eq:RMNz}
       R_{N-M}(z) & \equiv &  \prod_{i=M+1}^N (z - z_i) ~ .      
\eea
Since the roots of $Q_M(z)$ are known, it is easy to calculate explicitly the coefficients $y_k$ of the polynomial $Q_M(z)$, viz.
\be
     Q_M(z) = \sum_{k=0}^M y_k z^k
\ee
with
\bea
     y_M & = & 1 ~ , ~ \nonumber \\[2mm]
     y_{M-1} & = & - \sum_{m=1}^M z_m ~ , ~ \nonumber \\[2mm]
     y_{M-2} & = & + \sum_{m < m^\prime = 1}^M z_m z_{m^\prime}~ , ~ \nonumber \\[2mm]
     y_{M-3} & = & - \sum_{m < m^\prime < m^{\prime \prime} = 1}^M z_m z_{m^\prime} z_{m^{\prime \prime}} ~ , ~ \nonumber \\[2mm]
     \dots & & \nonumber \\[2mm]
     y_0 & = & (-)^M \prod_{m = 1}^M z_m ~ .
\eea

We now construct the modified correlator
\be
    \widetilde{C}_n^{(0)} \equiv  \sum_{k=0}^M y_k C_n^{(2k)} ~ , ~
    \label{eq:Cn_modified}
\ee
which implies
\bea
      \widetilde{C}_n^{(0)} & = & \sum_{i=1}^N A_i ~ \sum_{k=0}^M y_k z_i^k ~ \left[ e^{- a M_i n} + (-)^p e^{- a M_i (N_T - n)} \right] ~ , \nonumber \\[2mm]
                                       & = & \sum_{i=1}^N A_i ~ Q_M(z_i) ~ \left[ e^{- a M_i n} + (-)^p e^{- a M_i (N_T - n)} \right] ~ . ~
\eea
Since by definition $Q_M(z_i) = 0$ for $i = 1, ..., M$, the modified correlator does not contain any more the unwanted exponential signals, i.e.~those corresponding to the roots $z_i$ with $i = 1, ..., M$, while it contains only the remaining $N - M$ exponential signals
\be
     \widetilde{C}_n^{(0)} = \sum_{i=M+1}^{N-M} \widetilde{A}_i \left[ e^{- a M_i n} + (-)^p e^{- a M_i (N_T - n)} \right]
\ee
with
\be
     \widetilde{A}_i = A_i \cdot Q_M(z_i) ~ .
     \label{eq:Ai_modified}
\ee

We now apply the ODE algorithm to the modified correlator $\widetilde{C}_n^{(0)}$, check that it has $N-M$ roots at $z_i$ with $i = M+1, ..., N$ and obtain the amplitudes $\widetilde{A}_i$.
Finally we reconstruct the original amplitudes $A_i$ using Eq.~(\ref{eq:Ai_modified}).

The generalization to the case of roots with arbitrary multiplicities is straightforward.

We point out that the above filtering procedure can be very easily adapted when the masses of the unwanted signals are known {\it a priori} or from other analyses not based on the ODE algorithm.
In such cases the coefficients $y_k$ of the polynomial $Q_M(z)$ can be calculated directly without applying the ODE algorithm to the original correlator (\ref{eq:Cn_initial}).
Then, the filtering of the unwanted states can proceed by constructing the modified correlator (\ref{eq:Cn_modified}) and by applying to it the ODE algorithm.

\end{document}